\documentclass
[letterpaper,aps,prb,twocolumn,showpacs,nofootinbib,superscriptaddress]{revtex4}%
\usepackage{graphicx}
\usepackage{dcolumn}
\usepackage{bm}
\usepackage{amsmath}
\usepackage{amsfonts}
\usepackage{amssymb}%
\begin{document}
\title{Photon-Assisted Tunneling through Molecular Conduction Junctions with Graphene Electrodes.}
\author{Boris D. Fainberg}
\affiliation{Faculty of Science, Holon Institute of Technology, 5810201 Holon, Israel}
\affiliation{School of Chemistry, Tel-Aviv University, 69978 Tel-Aviv, Israel}
\date{\today }

\begin{abstract}
Graphene electrodes provide a suitable alternative to metal contacts in
molecular conduction nanojunctions. Here, we propose to use graphene
electrodes as a platform for effective photon assisted tunneling through
molecular conduction nanojunctions. We predict dramatic increasing currents
evaluated at side-band energies $\sim n\hbar\omega$ ($n$ is a whole number)
related to the modification of graphene gapless spectrum under the action of
external electromagnetic field of frequency $\omega$. A side benifit of using
doped graphene electrodes is the polarization control of photocurrent related
to the processes occurring either in the graphene electrodes or in the
molecular bridge. The latter processes are accompanied by surface plasmon
excitation in the graphene sheet that makes them more efficient. Our results
illustrate the potential of graphene contacts in coherent control of
photocurrent in molecular electronics, supporting the possibility of
single-molecule devices.

\end{abstract}

\maketitle

\section{Introduction}

The field of molecular-scale electronics has been rapidly advancing over the
past two decades, both in terms of experimental and numerical technology and
in terms of the discovery of new physical phenomena and realization of new
applications (for recent reviews please see
Refs.\cite{Kohler05,Chen09,Heath09}). In particular, the optical response of
nanoscale molecular junctions has been the topic of growing experimental and
theoretical interest in recent years
\cite{Park11PRB,Wang11PCCP,Fainberg_Galperin11PRB,Haertle10JCP,Reuter08PRL,Li08SSP,
Thanopulos08Nanotech,Prociuk08PRB,Li08NJP,Galperin08JCP,Li07EPL,Fai07PRB},
fueled in part by the rapid advance of the experimental technology and in part
by the premise for long range applications in optoelectronics.

A way of the control of the current through molecular conduction nanojunctions
is the well-known photon-assisted tunneling (PAT) \cite{Platero04,Kohler05}
that was studied already in the early 1960's experimentally by Dayem and
Martin \cite{Dayem_Martin62PRL} and theoretically by Tien and Gordon using a
simple theory which captures already the main physics of PAT
\cite{Tien_Gordon63}. The main idea is that an external field periodic in time
with frequency $\omega$ can induce inelastic tunneling events when the
electrons exchange energy quanta $\omega$ with the external field. PAT may be
related either to the potential difference modulation between the contacts of
the nanojunction when electric field is parallel to the axis of a junction
\cite{Tien_Gordon63,Gri98,Platero04,Kleinekathofer06EPL,Li07EPL}, or to the
electromagnetic (EM) excitation of electrons in the metallic contacts when
electric field is parallel to the film surface of contacts
\cite{Tien_Gordon63}. According to the Tien-Gordon model
\cite{Tien_Gordon63,Platero04,Li07EPL} for monochromatic external fields that
set up a potential difference $V(t)=V_{0}\cos\omega t$, the rectified dc
currents through ac-driven molecular junctions are determined as
\cite{Platero04,Li07EPL}%
\begin{equation}
I_{TG}=\sum_{n=-\infty}^{\infty}J_{n}^{2}(\frac{eV_{0}}{\hbar\omega}%
)I_{dc}^{0}(eV_{0}+n\hbar\omega)=\sum_{n=-\infty}^{\infty}I_{n} \label{eq:TG}%
\end{equation}
where the current in the driven system is expressed by a sum over
contributions of the current $I_{dc}^{0}(eV_{0}+n\hbar\omega)$ in\nolinebreak%
\ the\nolinebreak\ \nolinebreak undriven case but evaluated at side-band
energies $eV_{0}+n\hbar\omega$ shifted by integer multiples of the photon
quantum and weighted with squares of Bessel functions. A formula similar to
Eq.(\ref{eq:TG}) can be obtained also for EM excitation of electrons in the
metallic contacts \cite{Tien_Gordon63}. Note that the partial currents $I_{n}$
contain contributions from $\pm n$. The term $J_{n}(\frac{eV_{0}}{\hbar\omega
})$ denotes the $n$-th-order Bessel function of the first kind. The photon
absorption ($n>0$) and emission ($n<0$) processes can be viewed as creating
effective electron densities at energies $eV_{0}\pm n\hbar\omega$ with
probability $J_{n}^{2}(\frac{eV_{0}}{\hbar\omega})$. These probabilities
strongly diminish with number $n$ when $eV_{0}\leq\hbar\omega$ that severely
sidelines the control of the current for not strong EM fields ($<10^{6}$
$V/cm$ \cite{Kohler05}).

In the last time graphene, a single layer of graphite, with unusual
two-dimensional Dirac-like electronic excitations, has attracted considerable
attention due to its exceptional electronic properties (ballistic in-plane
electron transport etc.) \cite{Novoselov09RMP,Trauzettel07PRB,Efetov08PRB}.
Quite recently they have shown interest to a new kind of
graphene-molecule-graphene (GMG) junctions that may exhibit unique physical
properties, including a large conductance (achieving $0.38$ conductance
quantum), and are potentially useful as electronic and optoelectronic devices
\cite{Yang_graphene_junctions10JCP}. The junction consists of a conjugated
molecule connecting two parallel graphene sheets. In this relation it would be
interesting to investigate PAT\ in such a junction to control the current
through it. The PAT in GMG junctions under EM excitation of electrons and
holes in the graphene contacts may be rather different from that for usual
metallic contacts. It was shown that the massless energy spectrum of electrons
and holes in graphene led to the strongly non-linear EM response of this
system, which could work as a frequency multiplier \cite{Mikhailov07EPL}. The
predicted efficiency of the frequency up-conversion was rather high: the
amplitudes of the higher-harmonics of the ac electric current fell down slowly
(as $1/n$) with harmonics index $n$. Sure, the strongly non-linear EM response
should also lead to a slow falling down currents evaluated at side-band
energies $\sim n\hbar\omega$ (see Eq.(\ref{eq:TG})) with harmonics index $n$
in comparison to nanojunctions with metallic (or semiconductor
\cite{Fainberg13CPL}) leads (see below). This makes controlling charge
transfer essentially more effective than that for molecular nanojunctions with
metallic contacts. Additional factors that may enhance currents evaluated at
side-band energies $\sim n\hbar\omega$ in nanojunctions with graphene
electrodes are linear dependence of the density of states on energy in
graphene \cite{Novoselov09RMP}, and the gapless spectrum of graphene that can
change under the action of external EM field (see below).

Here we propose and explore theoretically a new approach to coherent control
of electric transport via molecular junctions, using either both graphene
electrodes or one graphene and another one - a metal electrode (that may be an
STM tip). Our approach is based on the excitation of dressed states of the
doped graphene electrode with electric field that is parallel to its surface,
having used unique properties of graphene mentioned above. As a first step, we
calculate a semiclassical wave function of a doped graphene under the action
of EM excitation. Then we obtain Heisenberg equations for the second
quantization operators of graphene and calculate current through a molecular
junction with graphene electrodes using non-equilibrium Green functions (GF).
We address different cases, which are analytically soluble, hence providing
useful insights. We show that using graphene electrodes can essentially
enhance currents evaluated at side-band energies $\sim n\hbar\omega$ in
molecular nanojunctions.

\section{Model Hamiltonian}

\begin{figure}
[ptb]
\begin{center}
\includegraphics[
height=4.4399in,
width=3.7645in
]%
{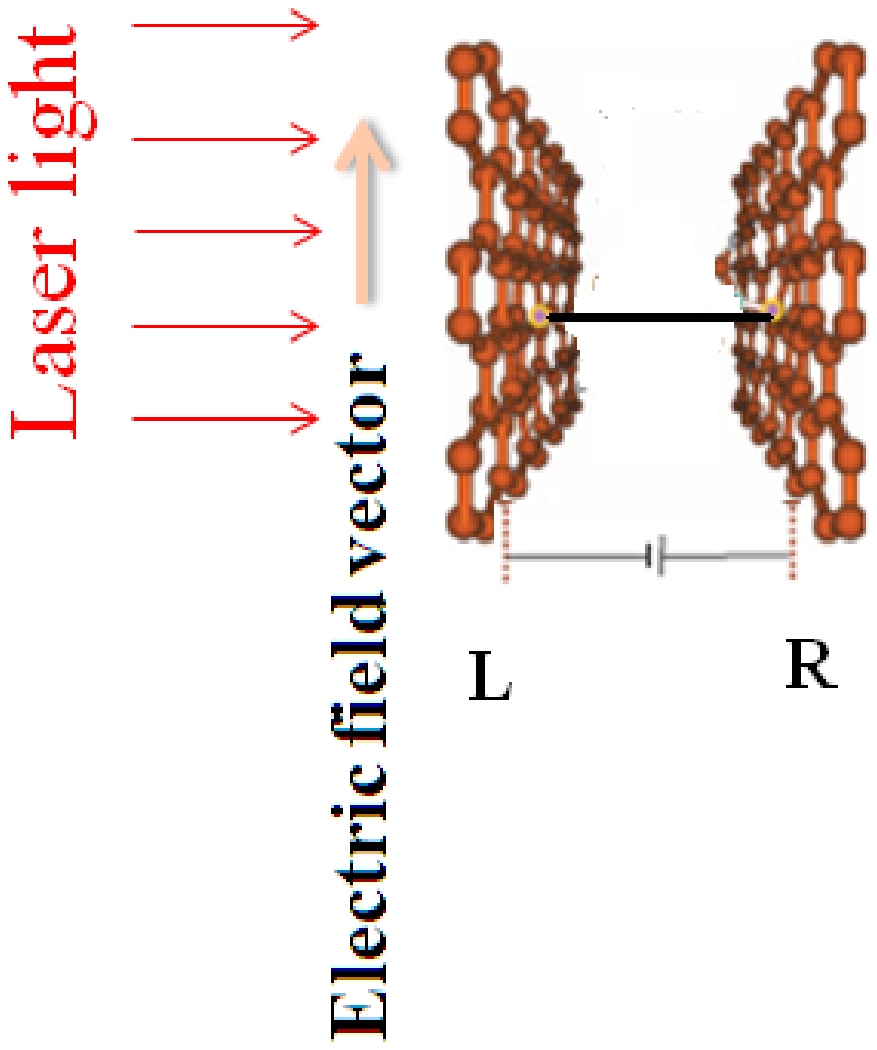}%
\caption{Molecular bridge ( thick horizontal line) between left (L) and right
(R) graphene electrodes with applied voltage bias. External electromagnetic
field acts on the electrodes.}%
\label{fig:GMG1}%
\end{center}
\end{figure}

\begin{figure}
[ptb]
\begin{center}
\includegraphics[
height=3.7308in,
width=2.6948in
]%
{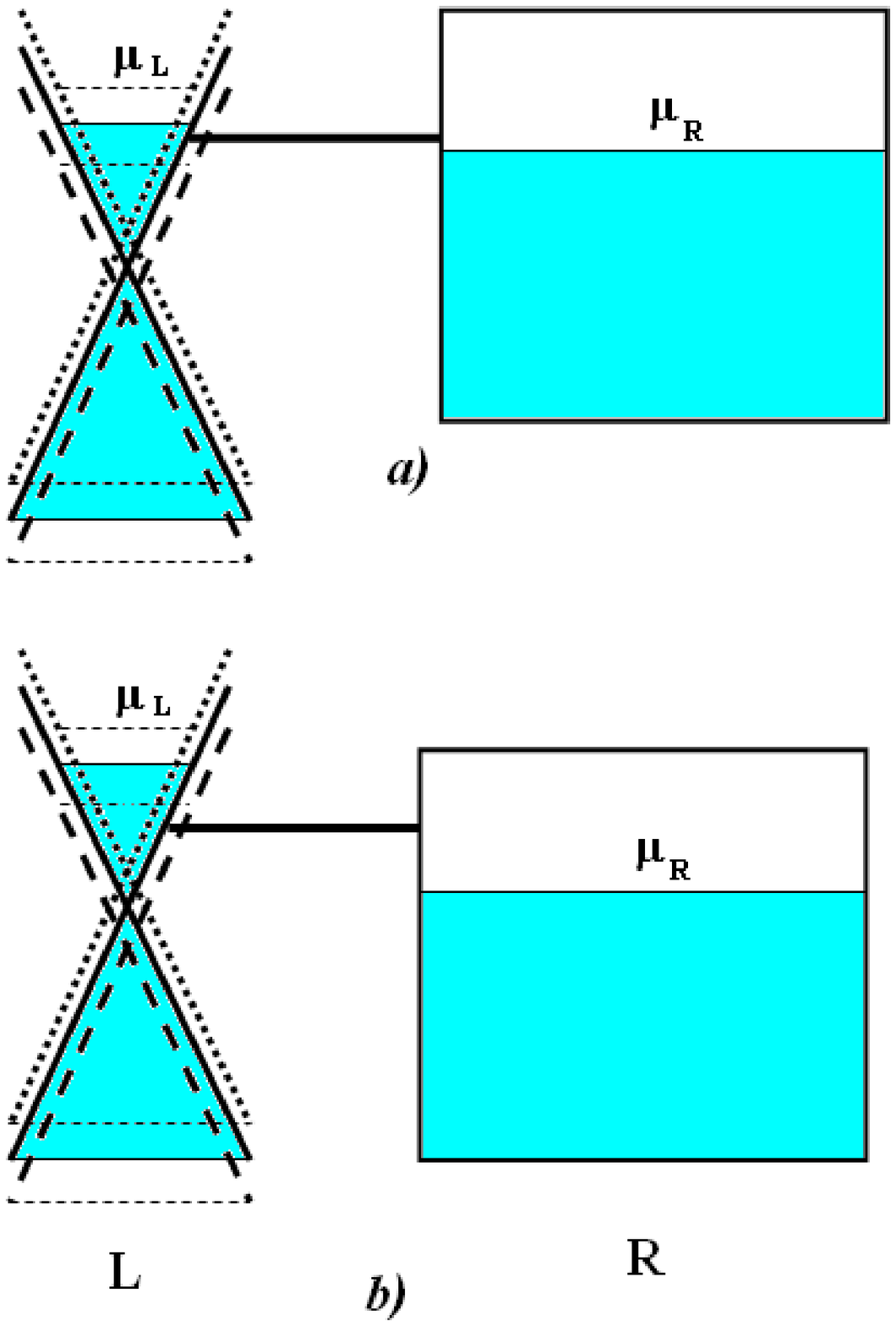}%
\caption{Molecular bridge between n-doped graphene (left-L) and metal
(right-R) electrodes. Thick horizontal line - energy of the molecular bridge
$\varepsilon_{m}$, $\mu_{L\text{ }}$ and $\mu_{R\text{ }}=\mu-e\varphi_{0}/2$
- chemical potentials of the left and right leads, respectively, in the biased
junction. The energy spectrum of unperturbated graphene is shown by the solid
line; dotted and dashed lines show the upper and lower first photonic replica
of the graphene spectrum, repectively, that are displaced an amount
$\hbar\omega$ from unperturbated spectrum. Solid thin horizontal line -
chemical potential of unperturbated graphene $\mu_{L\text{ }}=\mu+e\varphi
_{0}/2$, dashed thin horizontal lines - chemical potentials of the photonic
replica $\mu_{L\text{ }}=\mu+e\varphi_{0}/2\pm\hbar\omega$. \textit{a} -
potential of the graphene electrode is smaller than photon energy,
$e\varphi_{0}/2<\hbar\omega$; \textit{b} - potential of the graphene electrode
is larger than photon energy, $e\varphi_{0}/2>\hbar\omega$. The lower photonic
replication gives contribution into the current only in case \textit{b) } that
causes the step shown in Fig.\ref{fig:weak_field}.}%
\label{fig:photonic_replica}%
\end{center}
\end{figure}

Consider a spinless model for a molecular wire that comprises one site of
energy $\varepsilon_{m}$, positioned between either both graphene electrodes
(leads) (Fig.\ref{fig:GMG1}) or one graphene and another one - a metal
electrode (Fig.\ref{fig:photonic_replica}).%
%TCIMACRO{\FRAME{ftbpFU}{3.7645in}{4.4399in}{0pt}{\Qcb{Molecular bridge ( thick
%horizontal line) between left (L) and right (R) graphene electrodes with
%applied voltage bias. External electromagnetic field acts on the electrodes.}%
%}{\Qlb{fig:GMG1}}{gmg1.eps}{\special{ language "Scientific Word";
%type "GRAPHIC";  maintain-aspect-ratio TRUE;  display "USEDEF";
%valid_file "F";  width 3.7645in;  height 4.4399in;  depth 0pt;
%original-width 3.717in;  original-height 4.3889in;  cropleft "0";
%filename 'GMG1.eps';file-properties "XNPEU";}}}%
%BeginExpansion
%EndExpansion
The leads are represented by electron reservoirs $L$ and $R$, characterized by
the electronic chemical potentials $\mu_{K}$, $K=L,R$, and by the ambient
temperature $T$. The corresponding Fermi distributions are $f_{K}%
(\varepsilon_{k})=[\exp((\varepsilon_{k}-\mu_{K})/k_{B}T)+1]^{-1}$ in the
absense of external EM field, and the difference $\mu_{L}-\mu_{R}$
$=e\varphi_{0}$ is the imposed voltage bias between the electrodes. External
EM field acting on electrode $K$, $\mathbf{E}(t)=\mathbf{E}_{0}\cos\omega t$,
changes the corresponding Fermi distribution (see below). The Fermi energy of
the graphene electrode may be controlled via electrical or chemical
modification of the charge carrier density
\cite{Mak08PRL,Chen8Nature,Abajo11NL,Chen12Nature,Fei12Nature}. We consider
that steady-state current through a nanojunction does not influence on the
Fermi energy, since such current does not change a charge of the graphene electrode.%

%EndExpansion

The corresponding Hamiltonian is
\begin{equation}
\hat{H}_{junction}=\hat{H}_{wire}+\hat{H}_{leads}+\hat{V} \label{eq:H^}%
\end{equation}
where the wire Hamiltonian is $\hat{H}_{wire}=\varepsilon_{m}\hat{c}_{m}%
^{\dag}\hat{c}_{m}$, $\hat{c}_{m}^{\dag}$ ($\hat{c}_{m}$) are creation
(annihilation) operators for electrons at the molecular wire. The
molecule-leads interaction $\hat{V}$ describes electron transfer between the
molecular bridge and the right ($R$) and left ($L$) leads that gives rise to
net current in the biased junction
\begin{equation}
\hat{V}=\sum_{+,-}\sum_{\sigma,\mathbf{p}\in\{L,R\}}(V_{\mathbf{p\pm,}%
\sigma;m}\hat{a}_{\mathbf{p\pm,}\sigma}^{\dag}\hat{c}_{m}+H.c.)
\label{eq:V_M1}%
\end{equation}
Here $H.c.$ denotes Hermitian conjugate, $\hat{a}_{\mathbf{p\pm,}\sigma}%
^{\dag}$ are creation operators for graphene electrodes (see below). The
corresponding contribution to $\hat{V}$ from a metal electrode does not
contain summation with respect to positive and negative energies ($\pm$) and
quasispin index $\sigma$.

\section{Calculation of Semiclassical Wave Function}

The states of electrons in graphene are conveniently described by the
four-component wave functions, defined on two sublattices and two valleys.
Electron motion in the time-dependent EM field is described by the 2D Dirac
equation \cite{Novoselov09RMP,Efetov08PRB}
\begin{equation}
i\hbar\frac{\partial\mathbf{\psi}}{\partial t}=[v\mathbf{\hat{\sigma}%
}(\mathbf{\hat{p}}-\frac{e}{c}\mathbf{A})+e\varphi_{pot}]\mathbf{\psi}
\label{eq:psi_EFa}%
\end{equation}
written for a single valley and for a certain direction of spin. Here
$\mathbf{\hat{p}}$ is the momentum of the quasiparticle, $v$ - the Fermi
velocity ($v\approx10^{6}$ m/s), $\mathbf{\hat{\sigma}}$ - the vector of the
Pauli matrices in the sublattice space (\textquotedblleft
pseudospin\textquotedblright\ space), $\mathbf{A}$ and $\varphi_{pot}$ are
vector and scalar potentials of an EM field, respectively. Suppose a graphene
film is excited by a linearly polarized monochromatic electric field
$E_{x}(t)=E_{0}\cos\omega t$ that is parallel to its plane ($x,y$). Then
$A_{x}=-(c/\omega)E_{0}\sin\omega t$, $A_{y}=A_{z}=0$. Eq.(\ref{eq:psi_EFa})
can be brought to more symmetric form $i[\hat{P}-(e/c)\hat{A}]\mathbf{\psi
=}0,$ introducing matrices $\gamma_{1}=\hat{\sigma}_{y},\gamma_{2}%
=-\hat{\sigma}_{x}$ and $\gamma_{3}=\hat{\sigma}_{z}$, where%
\begin{equation}
\hat{P}=-i\hbar%
%TCIMACRO{\dsum \limits_{k=1}^{3}}%
%BeginExpansion
{\displaystyle\sum\limits_{k=1}^{3}}
%EndExpansion
\gamma_{k}\frac{\partial}{\partial x_{k}},\text{ }\hat{A}=%
%TCIMACRO{\dsum \limits_{k=1}^{3}}%
%BeginExpansion
{\displaystyle\sum\limits_{k=1}^{3}}
%EndExpansion
\gamma_{k}A_{x_{k}}, \label{eq:A^}%
\end{equation}
$x_{1}=x$, $x_{2}=y$, $x_{3}=ivt$ and $A_{x_{3}}=i\frac{c}{v}\varphi_{pot}$.
To obtain a semiclassical solution of Eq.(\ref{eq:psi_EFa}), we shall use a
method of Ref. \cite{Pauli32} (see also \cite{akhiezer-berestetskii69}). Let
us put $\mathbf{\psi}=-i(\hat{P}-\frac{e}{c}\hat{A})\Phi$. Then one can obtain
the following equation for $\Phi$%
\begin{equation}
\lbrack i\frac{\hbar e}{2c}%
%TCIMACRO{\dsum \limits_{k,l=1}^{3}}%
%BeginExpansion
{\displaystyle\sum\limits_{k,l=1}^{3}}
%EndExpansion
\gamma_{k}\gamma_{l}(1-\delta_{kl})F_{x_{l}x_{k}}-%
%TCIMACRO{\dsum \limits_{k=1\ }^{3}}%
%BeginExpansion
{\displaystyle\sum\limits_{k=1\ }^{3}}
%EndExpansion
(\hbar\frac{\partial}{\partial x_{k}}-i\frac{e}{c}A_{x_{k}})^{2}]\Phi=0
\label{eq:Fi_final}%
\end{equation}
where $F_{x_{l}x_{k}}=\partial A_{x_{l}}/\partial x_{k}-\partial A_{x_{k}%
}/\partial x_{l}$ is the field tensor. Let us seek a solution of
Eq.(\ref{eq:Fi_final}) as an expansion in power series in $\hbar$
\begin{equation}
\mathbf{\Phi=}\exp(iS/\hbar)w=\exp(iS/\hbar)(w_{0}+\hbar w_{1}+\hbar^{2}%
w_{2}+...) \label{eq:Fi_expansion}%
\end{equation}
where $S$ is a scalar and $w$ is a slowly varying spinor
\cite{berestetskii-lifshitz99}. Substituting series, Eq.(\ref{eq:Fi_expansion}%
), into Eq.(\ref{eq:Fi_final}) and collecting coefficients at the equal
exponents of $\hbar$, we get that $S$ is the action obeying the
Hamilton-Jacobi equation $\partial S/\partial t=-H$ where $H$ is the classical
Hamilton function of a particle:%
\begin{equation}
\exp(\frac{i}{\hbar}S)=\exp[-\frac{i}{\hbar}(v\int_{0}^{t}\sqrt{\bar{p}%
_{x}^{2}+\bar{p}_{y}^{2}}dt^{\prime}+e\int_{0}^{t}\varphi_{pot}dt^{\prime})],
\label{eq:exp(i/hS)}%
\end{equation}
and the equation for spinor $w_{0}$\begin{widetext}
\begin{eqnarray}
%\begin{equation}%
%TCIMACRO{\dsum \limits_{k=1\ }^{3}}%
%BeginExpansion
{\displaystyle\sum\limits_{k=1\ }^{3}}
%EndExpansion
\{[\frac{\partial}{\partial x_{k}}(\frac{\partial S}{\partial x_{k}}-\frac
{e}{c}A_{x_{k}})]w_{0}+2(\frac{\partial S}{\partial x_{k}}-\frac{e}{c}%
A_{x_{k}})\frac{\partial w_{0}}{\partial x_{k}}-\frac{e}{2c}%
%TCIMACRO{\dsum \limits_{l=1}^{3}}%
%BeginExpansion
{\displaystyle\sum\limits_{l=1}^{3}}
%EndExpansion
\gamma_{k}\gamma_{l}(1-\delta_{kl})F_{x_{l}x_{k}}w_{0}%
\}=0\label{eq:w_0_spinor3}%
%\end{equation}
\end{eqnarray}
\end{widetext}In Eq.(\ref{eq:exp(i/hS)}), $\mathbf{\bar{p}}$ is the normal
momentum that obeys the classical equations of motion $d\bar{p}_{x}%
/dt=-eE_{x}(t)$ for a particle with charge $-e$, according to which $\bar
{p}_{x}(t)=-(eE_{0}/\omega)\sin(\omega t)$; $\mathbf{\bar{p}=p}-\frac{e}%
{c}\mathbf{A}$ where $\mathbf{p}$ is the generalized momentum. If one takes
only the first term in series, Eq.(\ref{eq:Fi_expansion}), into account, it
can be shown that wave packets behave like particles moving along classical trajectories.

Let us solve Eq.(\ref{eq:w_0_spinor3}) for spinor $w_{0}$. We shall introduce
a linear combination of the components of the Hermitian conjugated wave
function $\mathbf{\psi}^{\dag}$ by $\mathbf{\bar{\psi}}=\mathbf{\psi}^{\dag
}\gamma_{3}$ \cite{akhiezer-berestetskii69}. Then using equation
$\mathbf{\psi}=-i(\hat{P}-\frac{e}{c}\hat{A})\Phi$ and Eqs.(\ref{eq:A^}), one
can show that electronic flux $s_{k}=i\mathbf{\bar{\psi}}\gamma_{k}%
\mathbf{\psi}$ obeys the continuity equation%
\begin{equation}%
%TCIMACRO{\dsum \limits_{k=1}^{3}}%
%BeginExpansion
{\displaystyle\sum\limits_{k=1}^{3}}
%EndExpansion
\frac{\partial}{\partial x_{k}}s_{k}=0 \label{eq:cont_eq}%
\end{equation}
Put
\begin{equation}
w_{0}=\sqrt{\xi}\varphi_{0} \label{eq:w_0}%
\end{equation}
where we denoted
\begin{equation}
\xi=-i2\bar{w}_{0}\hat{\pi}w_{0}%
\end{equation}
and $\hat{\pi}=%
%TCIMACRO{\dsum \limits_{k=1}^{3}}%
%BeginExpansion
{\displaystyle\sum\limits_{k=1}^{3}}
%EndExpansion
\gamma_{k}\pi_{k}$, $\pi_{k}=\partial S/\partial x_{k}-(e/c)A_{x_{k}}$. Then
in our approximation the electronic flux is reduced to $s_{k}=\pi_{k}\xi$ that
gives, bearing in mind Eq.(\ref{eq:cont_eq}),%

\begin{equation}%
%TCIMACRO{\dsum \limits_{k=1}^{3}}%
%BeginExpansion
{\displaystyle\sum\limits_{k=1}^{3}}
%EndExpansion
\frac{\partial}{\partial x_{k}}(\pi_{k}\xi)=0 \label{eq:cont_eq_SC}%
\end{equation}
Here quantities $\pi_{k}$ can be written as $\pi_{k}=\bar{p}_{k},$ $k=1,2$ and
$\pi_{3}=\pm i\bar{p}$ with the aid of the Hamilton-Jacobi equation $\partial
S/\partial t=-H$ and $\partial S/\partial x_{k}=p_{k},$ $k=1,2$. Here signs
plus and minus are related to positive and negative energies, respectively.
Eq.(\ref{eq:cont_eq_SC}) can be write over as%

\begin{equation}%
%TCIMACRO{\dsum \limits_{k=1\ }^{3}}%
%BeginExpansion
{\displaystyle\sum\limits_{k=1\ }^{3}}
%EndExpansion
(\frac{\partial\pi_{k}}{\partial x_{k}}\xi+\pi_{k}\frac{\partial\xi}{\partial
x_{k}})=0 \label{eq:cont_eq_SC2}%
\end{equation}
Using Hamilton's equations $\dot{x}_{k}=\partial H/\partial p_{k},$ $k=1,2,$
the time derivative $\dot{x}_{k}$ can be written as%
\begin{equation}
\dot{x}_{k}=\pm v\frac{\bar{p}_{k}}{\bar{p}}=iv\frac{\pi_{k}}{\pi_{3}},\text{
}k=1,2 \label{eq:x_k2}%
\end{equation}
This enables us to write down the second term on the right-hand side of
Eq.(\ref{eq:cont_eq_SC2}) in the form%

\[%
%TCIMACRO{\dsum \limits_{k=1\ }^{3}}%
%BeginExpansion
{\displaystyle\sum\limits_{k=1\ }^{3}}
%EndExpansion
\pi_{k}\frac{\partial\xi}{\partial x_{k}}=-\frac{i\pi_{3}}{v}[%
%TCIMACRO{\dsum \limits_{k=1\ }^{2}}%
%BeginExpansion
{\displaystyle\sum\limits_{k=1\ }^{2}}
%EndExpansion
\frac{\partial\xi}{\partial x_{k}}\frac{dx_{k}}{dt}+\frac{\partial\xi
}{\partial t}]=-\frac{i\pi_{3}}{v}\frac{d\xi}{dt}%
\]
and Eq.(\ref{eq:cont_eq_SC2}) becomes%
\begin{equation}
\frac{d\xi}{dt}=-\frac{1}{\bar{p}}\frac{\partial\bar{p}}{\partial t}%
\xi\label{eq:ksi}%
\end{equation}
since $\partial\pi_{k}/\partial x_{k}=0$ for $k=1,2$. Integrating
Eq.(\ref{eq:ksi}), one gets%
\begin{equation}
\xi(t)=\xi(0)\frac{\bar{p}(0)}{\bar{p}(t)} \label{eq:ksi(t)}%
\end{equation}
where $\bar{p}(0)=p$. Furthermore, substituting Eq.(\ref{eq:w_0}) into
Eq.(\ref{eq:w_0_spinor3}), we obtain equation for spinor $\varphi_{0}=\left(
\begin{array}
[c]{c}%
\varphi_{01}\\
\varphi_{02}%
\end{array}
\right)  $: $\frac{d\varphi_{0}}{dt}=\pm\frac{e}{2\bar{p}}\mathbf{\hat{\sigma
}E}\varphi_{0}$, the solution of which may be written as%
\begin{align}
\varphi_{01,2}  &  =\frac{1}{2\sqrt{p\bar{p}(1+\cos\varphi)(1+\cos\bar
{\varphi})}}\{\varphi_{01,2}(0)[\bar{p}(1+\cos\bar{\varphi})+\nonumber\\
&  +p(1+\cos\varphi)]\pm\varphi_{02,1}(0)[\bar{p}(1+\cos\bar{\varphi
})-p(1+\cos\varphi)]\} \label{eq:fi_01,2}%
\end{align}
The quantities $\xi(0)$ and $\varphi_{01,2}(0)$ in Eqs.(\ref{eq:ksi(t)}) and
(\ref{eq:fi_01,2}) are chosen in such a way that the wave function
$\mathbf{\psi}=\exp(\frac{i}{\hbar}S)(-i\hat{\pi})\sqrt{\xi}\varphi_{0}$
should be normalized and coincide with the wave function of unperturbated
graphene in the absence of external EM field \cite{Novoselov09RMP}. After
combersome calculations we get the wave function normalized for the graphene
sheet area $s$:%
\begin{align}
\psi &  =\frac{1}{\sqrt{s}}\exp(ip_{x}x/\hbar+ip_{y}y/\hbar)\exp[\frac
{i}{\hbar}(\mp v\int_{0}^{t}\bar{p}dt^{\prime}-\nonumber\\
&  -e\int_{0}^{t}\varphi_{pot}dt^{\prime})]\bar{u}_{\mathbf{p\pm}}
\label{eq:psi_norm2}%
\end{align}
where slowly varying spinors $\bar{u}_{\mathbf{p\pm}}$ are equal to
\begin{equation}
\bar{u}_{\mathbf{p\pm}}\mathbf{=}\frac{1}{\sqrt{2}}\left(
\begin{array}
[c]{c}%
\exp(-i\bar{\varphi}/2)\\
\pm\exp(i\bar{\varphi}/2)
\end{array}
\right)  , \label{eq:u^-NG}%
\end{equation}
$\bar{p}\equiv\left\vert \mathbf{\bar{p}}(t)\right\vert $, $\tan\bar{\varphi
}=\bar{p}_{y}/\bar{p}_{x}$, $p_{x}=p\cos\varphi$, $p_{y}=p\sin\varphi$,
$\tan\varphi=p_{y}/p_{x}$.

Eqs.(\ref{eq:psi_norm2}) and (\ref{eq:u^-NG}) show remarkable and very simple
result, according to which the time-dependent part of the semiclassical wave
function is defined by the same formula as that for the unperturbated system
with the only difference that the generalized momentum $\mathbf{p}$ should be
replaced by the usual momentum $\mathbf{\bar{p}}$. The space-dependent part of
the wave function remains unchanged.

\subsection{Heisenberg Equations for the Second Quantization Operators of
Graphene\textit{ }}

The wave function of the graphene sheet interacting with molecular bridge
$\Psi$ may be represented as the superposition of wave functions,
Eqs.(\ref{eq:psi_norm2}) and (\ref{eq:u^-NG}). Passing to the second
quantization, we get
\begin{align}
\Psi &  =\frac{1}{\sqrt{s}}%
%TCIMACRO{\dsum \limits_{+,-}}%
%BeginExpansion
{\displaystyle\sum\limits_{+,-}}
%EndExpansion%
%TCIMACRO{\dsum \limits_{\mathbf{p}}}%
%BeginExpansion
{\displaystyle\sum\limits_{\mathbf{p}}}
%EndExpansion
\hat{a}_{\mathbf{p\pm}}\exp[\frac{i}{\hbar}\mathbf{pr}+\frac{i}{\hbar}(\mp
v\int_{0}^{t}\bar{p}dt^{\prime}\nonumber\\
&  -e\int_{0}^{t}\varphi_{pot}dt^{\prime})]\bar{u}_{\mathbf{p\pm}}
\label{eq:Psi2}%
\end{align}
where $\hat{a}_{\mathbf{p\pm}}$ are annihilation operators. To obtain the
Hamiltonian in the second quantization representation, consider an average
energy of a particle with wave function $\psi$ that is given by $\int
\psi^{\ast}\hat{H}\psi d\mathbf{r=}i\hbar\int\psi^{\ast}(\partial\psi/\partial
t)d\mathbf{r}$. Replacing wave functions $\psi$ for $\Psi$ operators and
integrate with respect to $\mathbf{r}$, we get%
\begin{equation}
\hat{H}=\int\Psi^{\dag}\hat{H}\Psi d\mathbf{r}=%
%TCIMACRO{\dsum \limits_{\mathbf{p\sigma}}}%
%BeginExpansion
{\displaystyle\sum\limits_{\mathbf{p\sigma}}}
%EndExpansion%
%TCIMACRO{\dsum \limits_{+,-}}%
%BeginExpansion
{\displaystyle\sum\limits_{+,-}}
%EndExpansion
\hat{a}_{\mathbf{p\pm,}\sigma}^{\dag}\hat{a}_{\mathbf{p\pm,}\sigma}[\pm
v\bar{p}(t)+e\varphi_{pot}(t)] \label{eq:Hamiltonian_sc}%
\end{equation}
where $\sum_{\sigma}\hat{a}_{\mathbf{p\pm,}\sigma}^{\dag}\hat{a}%
_{\mathbf{p\pm,}\sigma}=\hat{a}_{\mathbf{p\pm}}^{\dag}\hat{a}_{\mathbf{p\pm}%
},$ $\sigma=1,2$ is the "quasispin" index. In deriving
Eq.(\ref{eq:Hamiltonian_sc}), we have taken into account that the main
contribution to $\partial\Psi/\partial t$ in the semiclassical approximation
is given by the exponential term on the right-hand side of Eq.(\ref{eq:Psi2})
(see Ref.\cite{landau-lifshitz.65}, chapter II). In addition, we beared in
mind that the summation over $\mathbf{p}$ can be substituted by the
integration over phase space $d\Gamma=d\mathbf{p}d\mathbf{r}$%
\begin{equation}%
%TCIMACRO{\dsum \limits_{\mathbf{p}}}%
%BeginExpansion
{\displaystyle\sum\limits_{\mathbf{p}}}
%EndExpansion
\rightarrow\int\frac{d\Gamma}{(2\pi\hbar)^{2}}=\frac{s}{(2\pi\hbar)^{2}}\int
d\mathbf{p} \label{eq:sum_p_to_integral}%
\end{equation}
Using Hamiltonian, Eq.(\ref{eq:Hamiltonian_sc}), we obtain the Heisenberg
equations of motion%

\begin{equation}
\frac{d\hat{a}_{\mathbf{p\pm,}\sigma}(t)}{dt}=\frac{i}{\hbar}[\hat{H},\hat
{a}_{\mathbf{p\pm,}\sigma}]\mathbf{\simeq}\frac{i}{\hbar}[\mp v\bar
{p}(t)-e\varphi_{pot}(t)]\hat{a}_{\mathbf{p\pm,}\sigma}(t) \label{eq:Heis}%
\end{equation}
\

\section{Formula for the Current}

The current from the $K$ lead ($K=L,R$) can be obtained by the generalization
of Eq.(12.11) of Ref.\cite{haug-jauho.96}%
\begin{equation}
I_{K}=-\frac{2\varkappa e}{\hbar}\operatorname{Re}\sum_{+,-}\sum
_{\sigma,\mathbf{p}\in K}V_{\mathbf{p\pm,}\sigma;m}G_{m;\mathbf{p\pm,}\sigma
}^{<}(t,t) \label{eq:I_K_GF}%
\end{equation}
where $\varkappa=1$ for the metal electrode, and $\varkappa=2$ for the
graphene electrode that accounts for the valley degeneracies of the
quasiparticle spectrum in graphene. $G_{m;\mathbf{p\pm,}\sigma}^{<}%
(t,t^{\prime})=i\langle\hat{a}_{\mathbf{p\pm,}\sigma}^{\dag}(t^{\prime}%
)\hat{c}_{m}(t)\rangle$ denotes the lesser GF that is given by
\begin{align}
G_{m;\mathbf{p\pm,}\sigma}^{<}(t,t^{\prime})  &  =\frac{1}{\hbar}\int
dt_{1}V_{\mathbf{p\pm,}\sigma;m}^{\ast}[G_{mm}^{r}(t,t_{1})g_{\mathbf{p\pm
,}\sigma}^{<}(t_{1},t^{\prime})+\nonumber\\
&  +G_{mm}^{<}(t,t_{1})g_{\mathbf{p\pm,}\sigma}^{a}(t_{1},t^{\prime})]
\label{eq:G<}%
\end{align}
where $G_{mm}^{r}(t,t_{1})$ and $G_{mm}^{<}(t,t_{1})$ are the retarded and
lesser wire GFs, respectively; $g_{\mathbf{p\pm,}\sigma}^{<}(t,t^{\prime
})=i\langle\hat{a}_{\mathbf{p\pm,}\sigma}^{\dag}(t^{\prime})\hat
{a}_{\mathbf{p\pm,}\sigma}(t)\rangle$ and $g_{\mathbf{p\pm,}\sigma}^{a}%
(t_{1},t^{\prime})=i\theta(t^{\prime}-t_{1})\langle\{\hat{a}_{\mathbf{p\pm
,}\sigma}(t_{1}),\hat{a}_{\mathbf{p\pm,}\sigma}^{\dag}(t^{\prime})\}\rangle$
are the lesser and advanced lead GFs, respectively; $\theta(t^{\prime}-t_{1})$
is the unit function. Using Eq.(\ref{eq:Heis}), we get
\begin{align}
g_{\mathbf{p\pm,}\sigma}^{<}(t,t^{\prime})  &  =i\langle\hat{a}_{\mathbf{p\pm
,}\sigma}^{\dag}(t^{\prime})\hat{a}_{\mathbf{p\pm,}\sigma}(t)\rangle
=if^{K}(vp_{\pm})\times\nonumber\\
&  \times\exp\{\frac{i}{\hbar}[-e\varphi_{pot,K}(t-t^{\prime})\mp
v\int_{t^{\prime}}^{t}dt^{\prime\prime}\bar{p}(t^{\prime\prime})]\}
\label{eq:g<2}%
\end{align}
and%

\begin{align}
g_{\mathbf{p\pm,}\sigma}^{a}(t_{1},t^{\prime})  &  =i\theta(t^{\prime}%
-t_{1})\exp\{\frac{i}{\hbar}[-e\varphi_{pot,K}(t_{1}-t^{\prime})\nonumber\\
&  \mp v\int_{t^{\prime}}^{t_{1}}dt^{\prime\prime}\bar{p}(t^{\prime\prime})]\}
\label{eq:g^a(t1,t')}%
\end{align}
where $f^{K}(vp_{\pm})$ $\equiv\langle\hat{a}_{\mathbf{p\pm,}\sigma}^{\dag
}(0)\hat{a}_{\mathbf{p\pm,}\sigma}(0)\rangle=\left[  1+\exp\left(  \frac{\pm
vp-\mu_{K}}{k_{B}T}\right)  \right]  ^{-1}$ is the Fermi function and $\mu
_{K}$ - the chemical potential of lead $K$. Substituting Eqs.(\ref{eq:G<}),
(\ref{eq:g<2}) and (\ref{eq:g^a(t1,t')}) into Eq.(\ref{eq:I_K_GF}), and
converting the momentum summations to energy integration,
Eq.(\ref{eq:sum_p_to_integral}), we get%

\begin{align}
I_{K}  &  =\frac{4e}{\hbar}\int_{-\infty}^{t}dt_{1}\sum_{+,-}\operatorname{Im}%
\int_{0}^{\infty}\frac{d(vp)}{2\pi}\exp[\pm\frac{i}{\hbar}e\varphi
_{pot,K}(t-t_{1})]\times\nonumber\\
&  \times\Gamma_{mm}^{K}(\pm vp,t_{1,}t)[G_{mm}^{r}(t,t_{1})f^{K}(\pm
vp)+G_{mm}^{<}(t,t_{1})] \label{eq:I_K_GF2}%
\end{align}
where%

\begin{align}
\Gamma_{mm}^{K}(\pm vp,t_{1,}t)  &  =\frac{2\pi}{\hbar}\left(  \frac{s}%
{2\pi^{2}\hbar v^{2}}\right)  \sum_{\sigma\in K}\int_{0}^{\pi}d\theta
vpV_{\mathbf{p\pm,}\sigma;m}(t)\times\nonumber\\
&  \times V_{\mathbf{p\pm,}\sigma;m}^{\ast}(t_{1})\exp[\pm\frac{i}{\hbar}%
v\int_{t_{1}}^{t}dt^{\prime}\bar{p}(t^{\prime})] \label{eq:Gamma^K}%
\end{align}
is the level-width function.

To proceed, we shall make the time expansion of $\Gamma_{mm}^{K}(\pm
vp,t_{1,}t)$ into the Fourier series, and then use the Markovian
approximation, considering time $t-t_{1}\equiv\tau$ as very short. This will
also enable us to use the non-interacting resonant-level model
\cite{haug-jauho.96} for finding the time dependence of $G_{mm}^{r}%
(t,t-\tau)=-i\theta(\tau)\exp(-\frac{i}{\hbar}\varepsilon_{m}\tau)$ and
$G_{mm}^{<}(t,t-\tau)=in_{m}(t)\exp(-\frac{i}{\hbar}\varepsilon_{m}\tau)$ as
functions of $t$ and $t-\tau$ where $n_{m}(t)$ is the population of molecular
state $m$.

According to the Floquet theorem \cite{Kohler05}, the general solution of the
Schr\"{o}dinger equation for an electron subjected to a periodic perturbation,
takes the form $\psi(t)=\exp(-\frac{i}{\hbar}\varepsilon t)\Phi_{T}(t)$, where
$\Phi_{T}(t)$ is a periodic function having the same period $T$ as the
perturbation, and $\varepsilon$ is called quasienergy. Then the expansion of
function $\exp[\frac{i}{\hbar}v\int_{0}^{t}dt^{\prime}\bar{p}(t^{\prime})]$ on
the right-hand side of Eq.(\ref{eq:psi_norm2}) into the Fourier series will be
as following%

\begin{equation}
\exp[\frac{i}{\hbar}v\int_{0}^{t}dt^{\prime}\bar{p}(t^{\prime})]=\exp[\frac
{i}{\hbar}\varepsilon(p,\theta)t]%
%TCIMACRO{\dsum \limits_{l=-\infty}^{\infty}}%
%BeginExpansion
{\displaystyle\sum\limits_{l=-\infty}^{\infty}}
%EndExpansion
c_{l}(p,\theta)\exp(ilt\omega) \label{eq:Fourier_expansion}%
\end{equation}
where%
\begin{equation}
c_{l}(p,\theta)=\frac{\omega}{2\pi}\int_{-\pi/\omega}^{\pi/\omega}\exp
[\frac{i}{\hbar}v\int_{0}^{t}dt^{\prime}\bar{p}(t^{\prime})-\frac{i}{\hbar
}\varepsilon(p,\theta)t-il\omega t]dt \label{eq:c_l(p,teta)}%
\end{equation}
Using expansion, Eq.(\ref{eq:Fourier_expansion}), into Eq.(\ref{eq:Gamma^K})
and neglecting fast oscillating with time $t$ terms, we get%

\begin{align}
\Gamma_{mm}^{K}(\pm vp,\tau)  &  =\frac{2\pi}{\hbar}\left(  \frac{s}{2\pi
^{2}\hbar v^{2}}\right)  \sum_{\sigma\in K}\int_{0}^{\pi}d\theta
vp|V_{\mathbf{p\pm,}\sigma;m}|^{2}\times\nonumber\\
&  \times%
%TCIMACRO{\dsum \limits_{n=-\infty}^{\infty}}%
%BeginExpansion
{\displaystyle\sum\limits_{n=-\infty}^{\infty}}
%EndExpansion
|c_{n}(p,\theta)|^{2}\exp\{\pm i[\frac{\varepsilon(p,\theta)}{\hbar}%
+n\omega]\tau\} \label{eq:Gamma_mm^K}%
\end{align}
Then going to the integration with respect to $\tau$ in Eq.(\ref{eq:I_K_GF2})
and bearing in mind Eq.(\ref{eq:Gamma_mm^K}), we get%
\begin{align}
I_{K}  &  =4e\sum_{\sigma\in K}\int_{0}^{\pi}d\theta%
%TCIMACRO{\dsum \limits_{n=-\infty}^{\infty}}%
%BeginExpansion
{\displaystyle\sum\limits_{n=-\infty}^{\infty}}
%EndExpansion
[n_{m}(t)-f^{K}(vp_{n\pm})]\times\nonumber\\
&  \times|c_{n}(p_{n\pm},\theta)|^{2}\bar{\gamma}_{G_{K}\sigma,m}^{(n)\pm}
\label{eq:I_Kgeneral2}%
\end{align}
where we denoted%
\begin{align}
\bar{\gamma}_{G_{K}\sigma,m}^{(n)\pm}  &  =\frac{s}{2\pi\hbar^{3}v^{2}}%
\int_{0}^{\infty}vpd(vp)|V_{\mathbf{p\pm,}\sigma;m}|^{2}\times\nonumber\\
&  \times\delta\lbrack\pm(\varepsilon(p,\theta)+n\hbar\omega)+e\varphi
_{pot,K}-\varepsilon_{m}] \label{eq:general_gamma^(n)+-}%
\end{align}
is the spectral function for the $n$-th photonic replication, $\delta(x)$ is
the Dirac delta, arguments $p_{n\pm}$ are defined by equation%
\begin{equation}
\varepsilon_{\pm}(p,\theta)=\pm(\varepsilon_{m}-e\varphi_{pot,K})-n\hbar
\omega\label{eq:epsilon(p)}%
\end{equation}
and should be positive. Below we shall consider $V_{\mathbf{p\pm,}\sigma;m}$
not dependent on $\mathbf{p\pm}$ and quasispin $\sigma$.

\section{Molecular Bridge between Graphene and Metal Electrodes}

Consider a specific case when the molecular bridge is found between graphene
and metal (tip) electrodes (Fig.\ref{fig:photonic_replica}). In that case one
can use Eq.(\ref{eq:I_Kgeneral2}) for $K=L$:
\begin{align}
I_{L}  &  =4e\sum_{\sigma\in K}%
%TCIMACRO{\dsum \limits_{n=-\infty}^{\infty}}%
%BeginExpansion
{\displaystyle\sum\limits_{n=-\infty}^{\infty}}
%EndExpansion
[n_{m}(t)-f^{L}(vp_{n\pm})]\times\nonumber\\
&  \times\int_{0}^{\pi}d\theta|c_{n}(p_{n\pm},\theta)|^{2}\bar{\gamma}%
_{G_{L}\sigma,m}^{(n)\pm} \label{eq:I7e}%
\end{align}
If $R$ represents the metal electrode, then%

\begin{equation}
I_{R}=2e\gamma_{Rm}[n_{m}(t)-f_{\mathbf{p}}^{R}] \label{eq:I_R1}%
\end{equation}
where $2\gamma_{Rm}$ is the charge transfer rate between the molecular bridge
and the metallic lead. In the case under consideration the equation
for\textit{ } $n_{m}(t)$ becomes%

\begin{equation}
\frac{dn_{m}}{dt}=-I_{L}/e-I_{R}/e \label{eq:n_m6}%
\end{equation}
that is written as the continuity equation. Inserting Eqs.(\ref{eq:I7e}) and
(\ref{eq:I_R1}) into Eq.(\ref{eq:n_m6}), solving the latter for the
steady-state regime and substituting the solution into Eq. (\ref{eq:I_R1}) for
current $I_{R}$, we get%

\begin{equation}
I_{R}=2e\gamma_{Rm}\frac{\sum_{\sigma}%
%TCIMACRO{\dsum \limits_{n=-\infty}^{\infty}}%
%BeginExpansion
{\displaystyle\sum\limits_{n=-\infty}^{\infty}}
%EndExpansion
\bar{\gamma}_{G_{L}\sigma,m}^{(n)\pm}\int_{0}^{\pi}d\theta|c_{n}(p_{n\pm
},\theta)|^{2}[f^{L}(vp_{n\pm})-f_{\mathbf{p}}^{R}]}{\sum_{\sigma}%
%TCIMACRO{\dsum \limits_{n=-\infty}^{\infty}}%
%BeginExpansion
{\displaystyle\sum\limits_{n=-\infty}^{\infty}}
%EndExpansion
\bar{\gamma}_{G_{L}\sigma,m}^{(n)\pm}\int_{0}^{\pi}d\theta|c_{n}(p_{n\pm
},\theta)|^{2}+\gamma_{Rm}/2} \label{eq:I_R2}%
\end{equation}
For a special case%
\[
\gamma_{Rm}/2>>\sum_{\sigma}%
%TCIMACRO{\dsum \limits_{n=-\infty}^{\infty}}%
%BeginExpansion
{\displaystyle\sum\limits_{n=-\infty}^{\infty}}
%EndExpansion
\bar{\gamma}_{G_{L}\sigma,m}^{(n)\pm}\int_{0}^{\pi}d\theta|c_{n}(p_{n\pm
},\theta)|^{2}%
\]
we obtain
\begin{equation}
I_{R}=4e\sum_{\sigma}%
%TCIMACRO{\dsum \limits_{n=-\infty}^{\infty}}%
%BeginExpansion
{\displaystyle\sum\limits_{n=-\infty}^{\infty}}
%EndExpansion
\int_{0}^{\pi}d\theta|c_{n}(p_{n\pm},\theta)|^{2}\bar{\gamma}_{G_{L}\sigma
,m}^{(n)\pm}[f^{L}(vp_{n\pm})-f_{\mathbf{p}}^{R}] \label{eq:I_R_special}%
\end{equation}
Eq.(\ref{eq:I_R_special}) seems similar to that of Tien and Gordon,
Eq.(\ref{eq:TG}), and generalizes it. To calculate current, we shall use a
variety of approaches.

\subsection{Calculations using Cumulant Expansions}

Function $\exp[\frac{i}{\hbar}v\int_{0}^{t}dt^{\prime}\bar{p}(t^{\prime})]$
may be written in the dimensionless form as%

\[
\exp(i\frac{\alpha}{b}\int_{0}^{y}dx\sqrt{1+2b\cos\theta\sin x+b^{2}\sin^{2}%
x})
\]
where $b\equiv(eE_{0}v/\omega)/(vp)$ and $\alpha=(eE_{0}v/\omega)/(\hbar
\omega)$ represent the work done by the electric field during one fourth of
period weighted per unperturbated energy $vp$ and photon energy $\hbar\omega$,
respectively; $y=\omega t$, and we assume $eE_{0}>0$. If $b<1$, one can use
the cumulant expansion, and we get%

\begin{align}
&  \exp[i\frac{\alpha}{b}\int_{0}^{y}dx\sqrt{1+2b\cos\theta\sin x+b^{2}%
\sin^{2}x}]\nonumber\\
&  =\exp[G_{1}(y)+G_{2}(y)]
\end{align}
where correct to fourth order with respect to $b$,%
\begin{align}
G_{1}(y)  &  =i\alpha\cos\theta(1-\frac{b^{2}}{3}\sin^{2}\theta)+i\frac
{\alpha}{b}[1+\frac{b^{2}}{4}\sin^{2}\theta-\nonumber\\
&  -\frac{3b^{4}}{64}\sin^{2}\theta(1-5\cos^{2}\theta)]y, \label{eq:G_1alpha}%
\end{align}

\begin{equation}
G_{2}(\tau)=iz_{1}\cos y+iz_{2}\sin2y+iz_{3}\cos3y+iz_{4}\sin4y
\label{eq:G_2z}%
\end{equation}
Here parameters $z_{l}\sim b^{l-1}$ are defined by $z_{1}=\alpha\cos
\theta\lbrack-1+(3/8)b^{2}\sin^{2}\theta],$ $z_{2}=(\alpha b/8)\sin^{2}%
\theta\lbrack-1+(b^{2}/4)(1-5\cos^{2}\theta)],$ $z_{3}=-(\alpha b^{2}%
/48)\sin2\theta\sin\theta$ and $z_{4}=-(\alpha b^{3}/256)\sin^{2}%
\theta(1-5\cos^{2}\theta)$.

As a matter of fact, the second term on the right-hand side of
Eq.(\ref{eq:G_1alpha}) that is proportional to $\tau$ describes the
quasienergy weight per photon energy%

\begin{equation}
\varepsilon(p,\theta)/(\hbar\omega)=\frac{\alpha}{b}[1+\frac{b^{2}}{4}\sin
^{2}\theta-\frac{3b^{4}}{64}\sin^{2}\theta(1-5\cos^{2}\theta)]
\end{equation}
that is anisotropic: $\varepsilon(p,\theta)=vp$ when the momentum is parallel
to electric field ($\theta=0$ or $\pi$), and is most different from $vp$ when
the momentum is perpendicular to the electric field ($\theta=\pi/2$). The term
$\exp[G_{2}(y)]$ can be expanded in terms of the Bessel functions $J_{s}%
(z_{i})$ as \cite{Abr64}%

\begin{align}
\exp(iz_{2n}\sin2ny)  &  =\sum_{s=-\infty}^{\infty}J_{s}(z_{2n})\exp
(i2sny),\nonumber\\
\exp[iz_{2n-1}\cos((2n-1)y)]  &  =\sum_{s=-\infty}^{\infty}J_{s}%
(z_{2n-1})\times\label{eq:expansion1}\\
&  \times\exp[is\frac{\pi}{2}+is(2n-1)y]\nonumber
\end{align}
where $n=1,2$. This gives expansion
\begin{align}
&  \left\vert c_{l}(p,\theta)\right\vert ^{2}\nonumber\\
&  =[\sum_{s_{2}s_{3}s_{4}}J_{l-2s_{2}-3s_{3}-4s_{4}}(z_{1})J_{-s_{2}}%
(z_{2})J_{-s_{3}}(z_{3})J_{s_{4}}(z_{4})]^{2} \label{eq:expansionBessel}%
\end{align}
for quantities $\left\vert c_{l}(p,\theta)\right\vert ^{2}$,
Eq.(\ref{eq:c_l(p,teta)}), that converge fast.

\begin{figure}
[ptb]
\begin{center}
\includegraphics[
height=3.4359in,
width=3.6703in
]%
{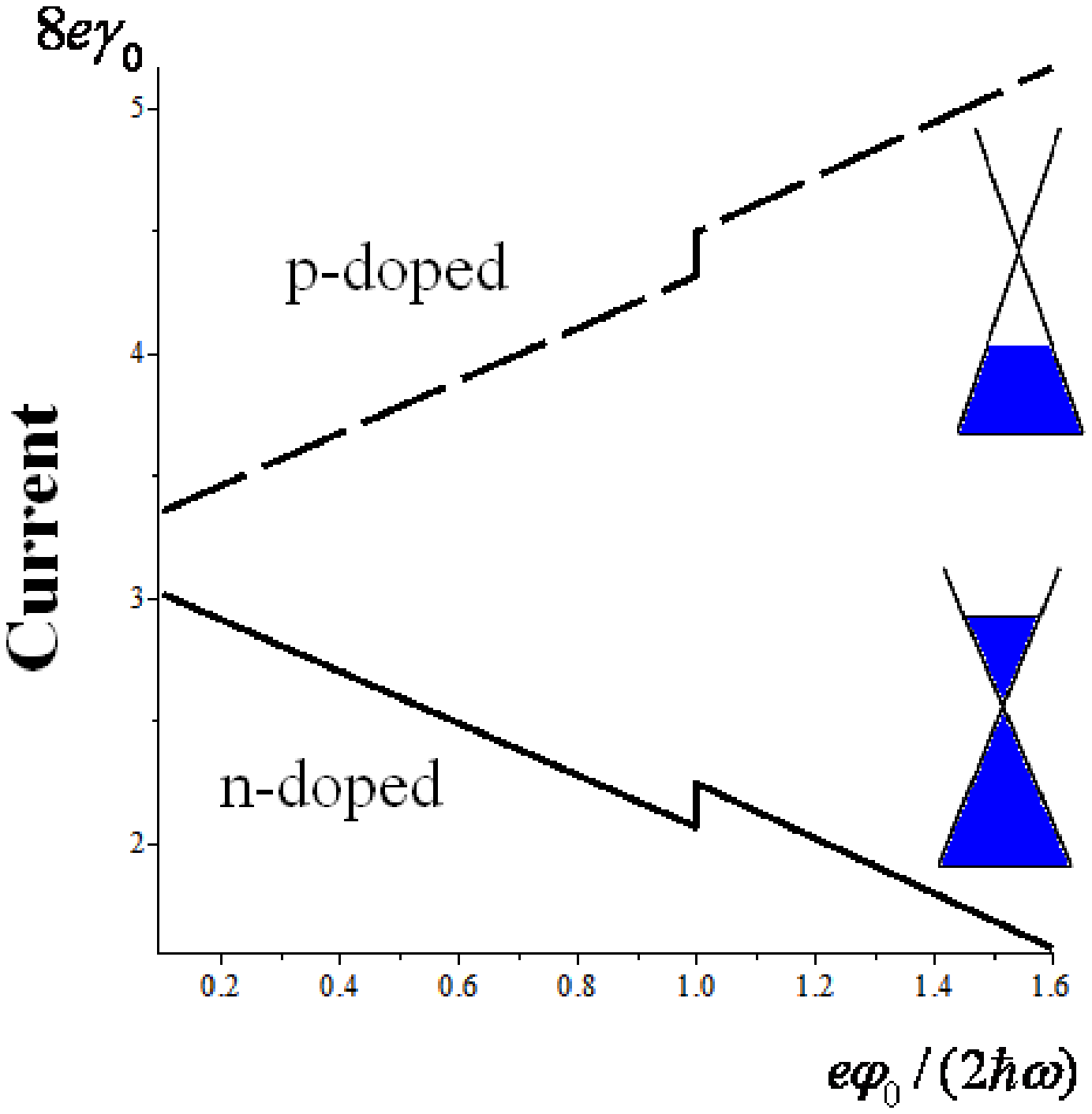}%
\caption{Current in the linear regime for n-doped ($\mu>0$, solid) and p-doped
($\mu<0$, dashed) graphene electrode as a function of applied voltage bias.
$|\varepsilon_{m}|=3\hbar\omega$, $\alpha=0.7$.}%
\label{fig:weak_field}%
\end{center}
\end{figure}

For a linear case (weak fields) $\left\vert c_{0}(p,\theta)\right\vert
^{2}\approx1$, $\left\vert c_{\pm1}(p,\theta)\right\vert ^{2}\approx
(\alpha\cos\theta)^{2}/4$, $\varepsilon(p,\theta)\approx vp$, and we get from
Eq.(\ref{eq:epsilon(p)}): $vp_{n\pm}=\pm(\varepsilon_{m}-e\varphi
_{pot,K})-n\hbar\omega$. In that case quantities $\bar{\gamma}_{G_{L}\sigma
,m}^{(n)\pm}$, Eq.(\ref{eq:general_gamma^(n)+-}), become%
\begin{equation}
\bar{\gamma}_{G_{L}\sigma,m}^{(n)\pm}=\frac{\gamma_{0}}{\pi}[\pm
\frac{(\varepsilon_{m}-e\varphi_{pot,L})}{\hbar\omega}-n]
\label{eq:new_gamma^(n)+-}%
\end{equation}
where $\gamma_{0}=|V_{\mathbf{p\pm,}\sigma;m}|^{2}s\omega/(2\hbar^{2}v^{2})$,
and the expression in the square brackets is proportional to the DOS for
graphene that is proportional to energy \cite{Novoselov09RMP}. The current,
Eq.(\ref{eq:I_R_special}), calculated in the linear regime using
Eq.(\ref{eq:new_gamma^(n)+-}), as a function of applied voltage bias is shown
in Fig.\ref{fig:weak_field}. In our calculations temperature $T=0$, and the
leads chemical potentials in the biased junction were taken to align
symmetrically with respect to the energy level $\varepsilon_{m}$
\cite{Li_Fai12Nano_Let}, i.e., $\mu+e\varphi_{0}/2$ for the left lead, and
$\mu-e\varphi_{0}/2$ for the right lead ($e\varphi_{0}\geq0$, $e\varphi
_{pot,(L,R)}=\pm e\varphi_{0}/2$) where $\mu=\varepsilon_{m}$ for both leads.
Both curves of Fig.\ref{fig:weak_field} show photon assisted current -
%TCIMACRO{\FRAME{ftbpFU}{3.6703in}{3.4359in}{0pt}{\Qcb{Current in the linear
%regime for n-doped ($\mu>0$, solid) and p-doped ($\mu<0$, dashed) graphene
%electrode as a function of applied voltage bias. $|\varepsilon_{m}%
%|=3\hbar\omega$, $\alpha=0.7$.}}{\Qlb{fig:weak_field}}{irweak_n+p2.eps}%
%width 3.6703in;  height 3.4359in;  depth 0pt;  original-width 6.0701in;
%original-height 5.6818in;  cropleft "0";  croptop "1";  cropright "1";
%cropbottom "0";  filename 'IRweak_n+p2.eps';file-properties "XNPEU";}}}%
%BeginExpansion
%EndExpansion
the steps when the potential of the graphene electrode achieves the value
corresponding to the photon energy. The steps are found on the background that
decreases linearly for a n-doped graphene electrode and increases linearly for
a p-doped electrode when $e\varphi_{0}$ increases. This is related to the
linear dependence of DOS on energy. Fig.\ref{fig:photonic_replica} shows our
model together with the photonic replica of the graphene electrodes and
elucidates the behavior observed in Fig.\ref{fig:weak_field}.

\begin{figure}
[ptb]
\begin{center}
\includegraphics[
height=3.3105in,
width=3.4151in
]%
{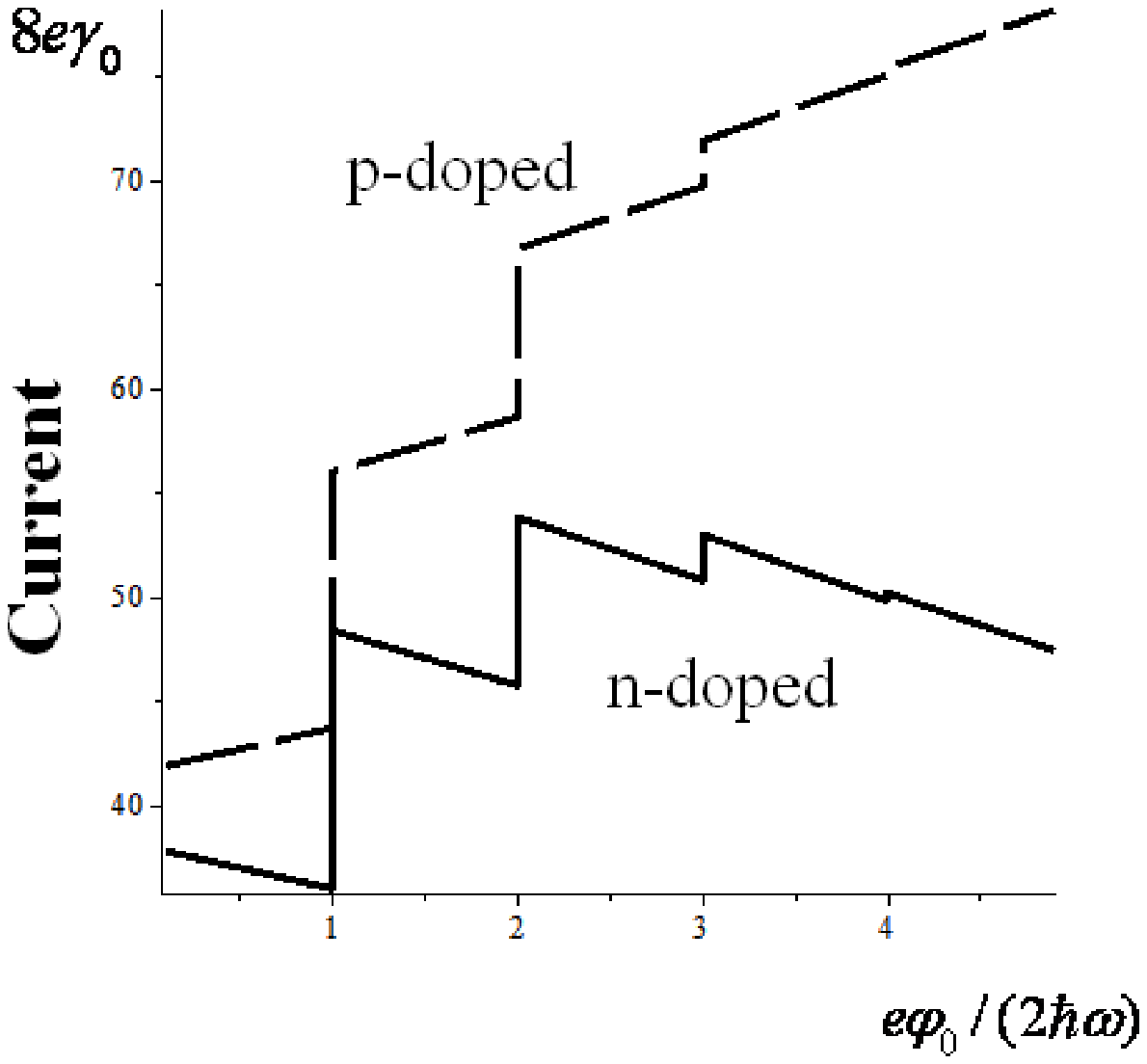}%
\caption{Current in the case of large momenta for n-doped ($\mu>0$, solid) and
p-doped ($\mu<0$, dashed) graphene electrode as a function of applied voltage
bias. $|\varepsilon_{m}|=20\hbar\omega$, $\alpha=3$.}%
\label{fig:large_momentum_n+p-doped}%
\end{center}
\end{figure}

When the interaction with external field is not small, $\alpha\geq1$, the
linear consideration does not apply. In case of large momenta (far from the
Dirac point), $b<<1$, Eq.(\ref{eq:new_gamma^(n)+-}) applies, and we get from
Eq.(\ref{eq:expansionBessel}) $\left\vert c_{l}(p,\theta)\right\vert
^{2}=J_{l}^{2}(\alpha\cos\theta)$. The current, Eq.(\ref{eq:I_R_special}),
calculated for large momenta when $\alpha=3$, as a function of applied voltage
bias is shown in Fig.\ref{fig:large_momentum_n+p-doped}. The number of steps
and their heights increase in comparison with the linear case.%

%TCIMACRO{\FRAME{ftbpFU}{3.4151in}{3.3105in}{0pt}{\Qcb{Current in the case of
%large momenta for n-doped ($\mu>0$, solid) and p-doped ($\mu<0$, dashed)
%graphene electrode as a function of applied voltage bias. $|\varepsilon
%_{m}|=20\hbar\omega$, $\alpha=3$.}}{\Qlb{fig:large_momentum_n+p-doped}%
%}{irlarge_p_n+p-doped.eps}{\special{ language "Scientific Word";
%type "GRAPHIC";  maintain-aspect-ratio TRUE;  display "USEDEF";
%valid_file "F";  width 3.4151in;  height 3.3105in;  depth 0pt;
%original-width 5.0557in;  original-height 4.9in;  cropleft "0";  croptop "1";
%cropright "1";  cropbottom "0";
%filename 'IRlarge_p_n+p-doped.eps';file-properties "XNPEU";}}}%
%BeginExpansion
%EndExpansion

\subsection{Calculations of Current including Small Momenta}

To calculate coefficients $c_{l}(p,\theta)$, Eq.(\ref{eq:c_l(p,teta)}), in
general case, we need to know quasienergy $\varepsilon(p,\theta)$. The latter
may be found as zero harmonic of the Fourier cosine series of normal momentum
$\bar{p}(t)$ on the left-hand side of Eq.(\ref{eq:c_l(p,teta)}). Consider
first limiting points $\theta=0,\pi$ when the momentum is parallel to the
electric field. Then the quasienergy weighted per the work done by the
electric field during one fourth of period is equal to $\bar{\varepsilon
}(p;\theta=0,\pi)\equiv\varepsilon(p;\theta=0,\pi)/(evE_{0}/\omega)=[1/(2\pi
b)]\int_{-\pi}^{\pi}dx\left\vert 1\pm b\sin x\right\vert $. If $b<1$,
$\bar{\varepsilon}(p;\theta=0,\pi)=1/b\sim vp$ like above. When $b>1$,
\begin{equation}
\bar{\varepsilon}(p;\theta=0,\pi)=\frac{2}{\pi b}[\arcsin(\frac{1}{b}%
)+\sqrt{1-\frac{1}{b^{2}}}] \label{eq:epsilon(p,teta=0,Pi)}%
\end{equation}
that gives for $b>>1$
\begin{equation}
\varepsilon(p;\theta=0,\pi)=\frac{1}{\pi}[2\alpha\hbar\omega+\frac{(vp)^{2}%
}{evE_{0}/\omega}] \label{eq:epsilon(p,teta=0,Pi;b>>1)}%
\end{equation}
- a quadratic dependence of $\varepsilon(p;\theta=0,\pi)$ on $vp$ for small
$vp$ or large $evE_{0}/\omega$ accompanied by opening the gap $4\alpha
\frac{\hbar\omega}{\pi}$ (see Fig.\ref{fig:quasienergy} below). This gap is
different from those predicted in Refs.\cite{Efetov08PRB},
\cite{Oka_Aoki09PRB}, which are induced by interband transitions in an undoped
graphene. In contrast, a semiclassical approximation used in our work is
correct for doped graphene when $\hbar\omega<2\mu$ \cite{Mikhailov07EPL}, and
as a consequence, interband transitions are excluded. Therefore, in our case
the gap is induced by intraband processes. When $\varepsilon(p;\theta=0,\pi)$
is defined by Eq.(\ref{eq:epsilon(p,teta=0,Pi;b>>1)}), quantities $\bar
{\gamma}_{G_{L}\sigma,m}^{(n)\pm},$ Eq.(\ref{eq:general_gamma^(n)+-}), become
$\bar{\gamma}_{G_{L}\sigma,m}^{(n)\pm}=\alpha\gamma_{0}/4$ that do not depend
on $n$ and are proportional to $\alpha$.%

%TCIMACRO{\FRAME{ftbpFU}{3.5379in}{2.8928in}{0pt}{\Qcb{The logarithm of the
%absolute values of Fourier-coefficients $c_{l}(p;\theta=0,\pi)$ (solid line)
%versus harmonic number $l$ for n-doped graphene contact ($\mu>0$) and
%$\alpha=0.5$, $b=1.43>1$. For comparison we also show $|J_{l}(\alpha)|$
%(dashed line). We use the continuous variable $l$ though $l$ takes only the
%whole values.}}{\Qlb{fig:abs(cl),theta=0}}{abs(cl).eps}%
%width 3.5379in;  height 2.8928in;  depth 0pt;  original-width 5.015in;
%original-height 4.0932in;  cropleft "0";  croptop "1";  cropright "1";
%cropbottom "0";  filename 'abs(cl).eps';file-properties "XNPEU";}}}%
%BeginExpansion
\begin{figure}
[ptb]
\begin{center}
\includegraphics[
height=2.8928in,
width=3.5379in
]%
{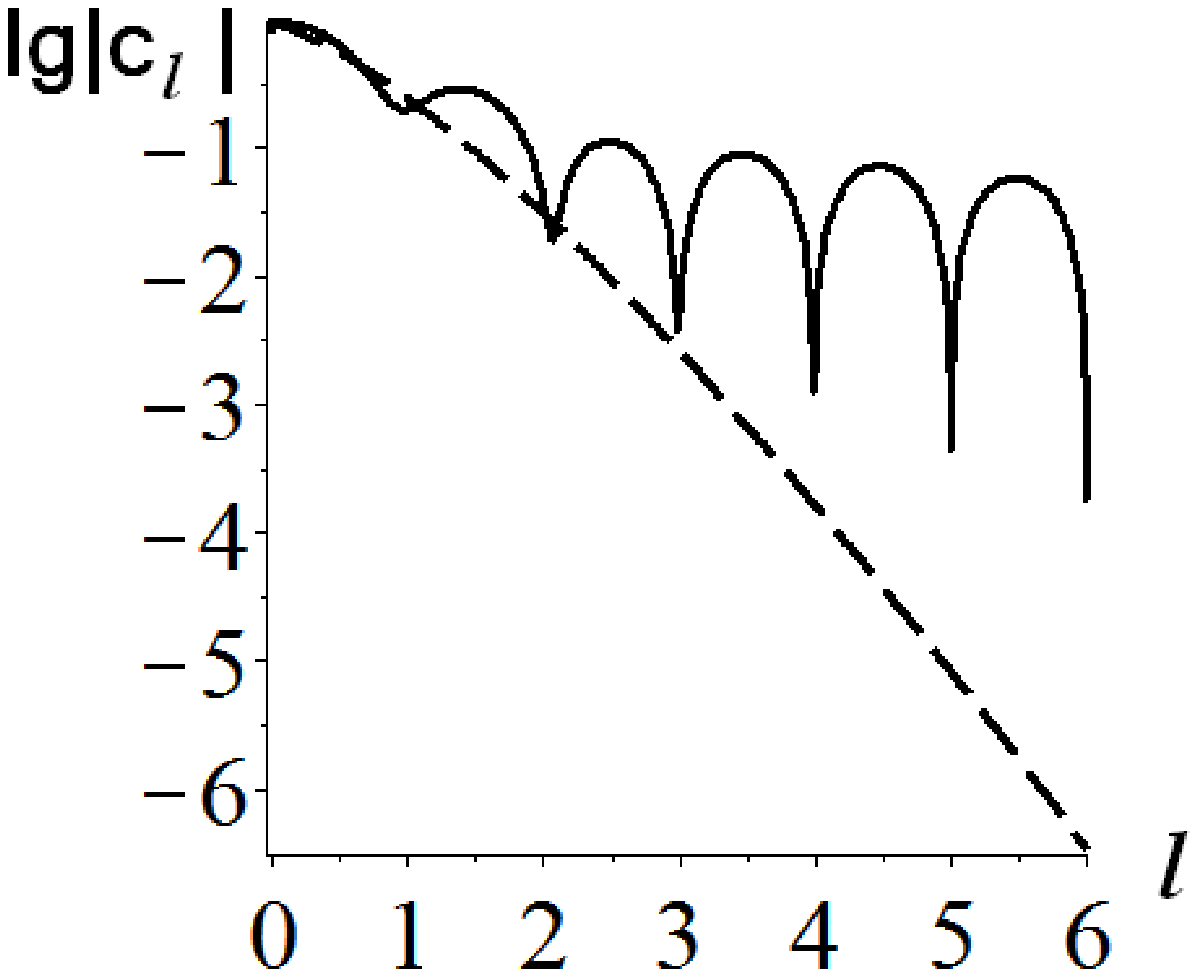}%
\caption{The logarithm of the absolute values of Fourier-coefficients
$c_{l}(p;\theta=0,\pi)$ (solid line) versus harmonic number $l$ for n-doped
graphene contact ($\mu>0$) and $\alpha=0.5$, $b=1.43>1$. For comparison we
also show $|J_{l}(\alpha)|$ (dashed line). We use the continuous variable $l$
though $l$ takes only the whole values.}%
\label{fig:abs(cl),theta=0}%
\end{center}
\end{figure}
%EndExpansion
Fig.\ref{fig:abs(cl),theta=0} shows the logarithm of the absolute values of
Fourier-coefficients $c_{l}^{+}(p;\theta=0,\pi)$ for different $l$ calculated
using Eqs.(\ref{eq:c_l(p,teta)}), (\ref{eq:epsilon(p)}) and
(\ref{eq:epsilon(p,teta=0,Pi)}). For comparison we also show the usual
dependence $\left\vert c_{l}(p;\theta=0,\pi)\right\vert =|J_{l}(\alpha)|$. One
can see much slower falling down $\left\vert c_{l}^{+}(p;\theta=0,\pi
)\right\vert $ with harmonics index $l$ in comparison to the usual dependence
that may be explained by the peculiarities of the graphene spectrum.

One can show that $\left\vert c_{l}(p,\theta\right\vert $ falls down as $1/l$
for $b>>1$ and $\alpha<<1$. Indeed, using
Eqs.(\ref{eq:epsilon(p,teta=0,Pi;b>>1)}) and
(\ref{eq:epsilon(p,teta=Pi/2;b>>1)}), one can obtain for Fourier-coefficients
$c_{l}(p,\theta)$, Eq.(\ref{eq:c_l(p,teta)}),
\begin{equation}
c_{l}(p,\theta)=\frac{1}{\pi}\operatorname{Re}%
%TCIMACRO{\dint \limits_{0}^{\pi}}%
%BeginExpansion
{\displaystyle\int\limits_{0}^{\pi}}
%EndExpansion
\exp[i\alpha(\cos\tau-1)+i\tau(l+\frac{2\alpha}{\pi})]d\tau\label{eq:c_l}%
\end{equation}
when $b>>1$ (small momenta). To calculate integral on the right-hand side of
Eq.(\ref{eq:c_l}), we use expansion, Eq.(\ref{eq:expansion1}), that gives%
\begin{equation}
c_{l}(p,\theta)=\frac{2}{\pi}\sum_{n=-\infty}^{\infty}\frac{J_{n}(\alpha
)}{l+\frac{2\alpha}{\pi}+n}\left\{
\begin{array}
[c]{c}%
(-1)^{n/2}\sin\alpha\text{, }n\text{ is even}\\
(-1)^{\frac{n+1}{2}+1}\cos\alpha\text{, }n\text{ is odd }%
\end{array}
\right\}  \label{eq:c_l2}%
\end{equation}
where $l=2k$ is even. If $l=2k+1$, $c_{l}(p,\theta)=0$. Eq.(\ref{eq:c_l2})
gives $c_{0}(p,\theta)\simeq1$ and%

\begin{equation}
c_{l}(p,\theta)\simeq\frac{2\alpha}{\pi}(\frac{1}{l}+\frac{l}{l^{2}-1}),\text{
}l\geqslant2 \label{eq:c_l3}%
\end{equation}
for $\alpha<<1$. Eq.(\ref{eq:c_l3}) shows that $c_{l}(p,\theta)\sim1/l$ for
$l>2$. Such a behaviour is due to stronly non-linear EM response of graphene,
which could also work as a frequency multiplier \cite{Mikhailov07EPL}. Our
approach enables us to understand the origin of this non-linear response that
arises due to modification of graphene gapless spectrum in the external EM field.

Consider now the middle point $\theta=\pi/2$ when the momentum is
perpendicular to the electric field. In that case one can show that
\begin{align}
\bar{\varepsilon}(p;\theta &  =\pi/2)=\frac{1}{2\pi b}\int_{-\pi}^{\pi}%
dx\sqrt{1+b^{2}\sin^{2}x}=\nonumber\\
&  =\frac{2}{\pi}\sqrt{1+b^{-2}}E[(1+b^{-2})^{-1/2}]
\label{eq:epsilon(p,teta=Pi/2)}%
\end{align}
where $E(x)$ is the complete elliptic integral of the second kind
\cite{Abr64}. If $b\ll1$, $\bar{\varepsilon}(p,\pi/2)=1/b$ like before. When
$b>>1$, we get
\begin{equation}
\varepsilon(p,\theta=\frac{\pi}{2})=\frac{1}{\pi}\{2\alpha\hbar\omega
+[\frac{1}{2}+2\ln(2\sqrt{\frac{eE_{0}}{\omega p}})]\frac{(vp)^{2}}%
{evE_{0}/\omega}\} \label{eq:epsilon(p,teta=Pi/2;b>>1)}%
\end{equation}
where the dependence of $\varepsilon(p,\pi/2)$ on $p$ for small $p$ (or large
$eE_{0}/v)$ differs from quadratic one (cf. with
Eq.(\ref{eq:epsilon(p,teta=0,Pi;b>>1)})). Hence, the quasienergy becomes
anisotropic, however, its formation is accompanied by opening the same
dynamical gap $4\alpha\frac{\hbar\omega}{\pi}$ as for $\theta=0,\pi$.
Quasienergies $\bar{\varepsilon}(p;\theta=0,\pi,\pi/2)$ defined by
Eqs.(\ref{eq:epsilon(p,teta=0,Pi)}) and (\ref{eq:epsilon(p,teta=Pi/2)}) as
functions of $1/b=vp/(eE_{0}v/\omega)$ are shown in Fig.\ref{fig:quasienergy}.
They are equal to $2/\pi$ for zero momentum, then increase as $\sim(vp)^{2}$
for $\theta=0,\pi$, Eq.(\ref{eq:epsilon(p,teta=0,Pi;b>>1)}), and according to
Eq.(\ref{eq:epsilon(p,teta=Pi/2;b>>1)}) for $\theta=\pi/2$. The law,
Eq.(\ref{eq:epsilon(p,teta=0,Pi)}), for $\theta=0,\pi$ gives way to linear one
when $1/b=1$, and quasienergy for $\theta=\pi/2$ also tends to linear one when
$1/b>>1$ (large momenta).%

%TCIMACRO{\FRAME{ftbpFU}{3.4566in}{3.397in}{0pt}{\Qcb{Quasienergies
%$\bar{\varepsilon}(p;\theta)$ for $\theta=0,\pi$ (solid line) and $\pi/2$
%(dashed line) as functions of $1/b=p\omega/(eE_{0})$.}}{\Qlb{fig:quasienergy}%
%}{quasienergy.eps}{\special{ language "Scientific Word";  type "GRAPHIC";
%width 3.4566in;  height 3.397in;  depth 0pt;  original-width 5.0427in;
%original-height 4.9545in;  cropleft "0";  croptop "1";  cropright "1";
%cropbottom "0";  filename 'quasienergy.eps';file-properties "XNPEU";}}}%
%BeginExpansion
\begin{figure}
[ptb]
\begin{center}
\includegraphics[
height=3.397in,
width=3.4566in
]%
{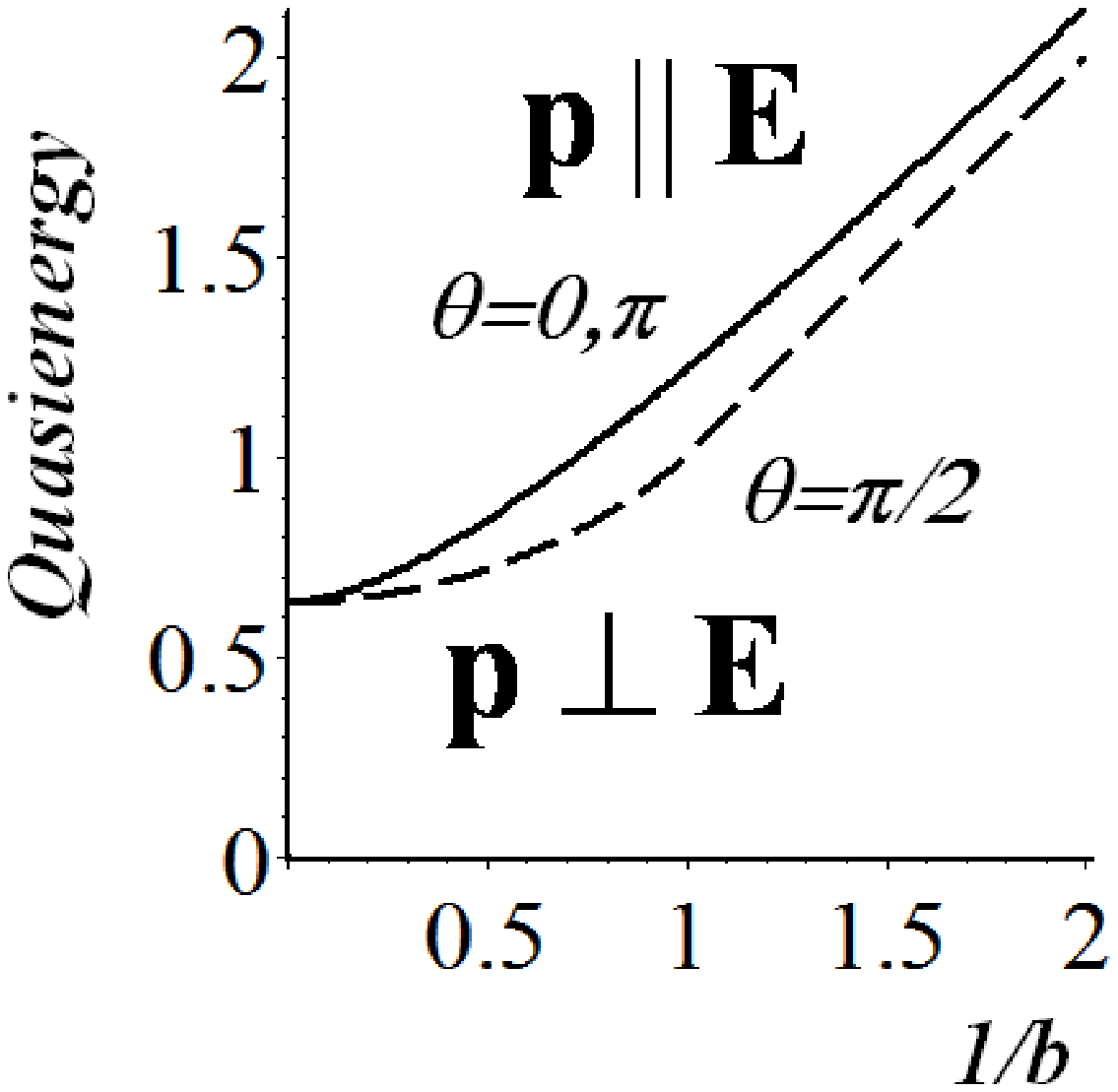}%
\caption{Quasienergies $\bar{\varepsilon}(p;\theta)$ for $\theta=0,\pi$ (solid
line) and $\pi/2$ (dashed line) as functions of $1/b=p\omega/(eE_{0})$.}%
\label{fig:quasienergy}%
\end{center}
\end{figure}
%EndExpansion

\section{Conclusion and Outlook}

Here we have proposed and explored theoretically a new approach to coherent
control of electric transport via molecular junctions, using graphene
electrodes. Our approach is based on the excitation of dressed states of the
doped graphene with electric field that is parallel to its surface, having
used unique properties of the graphene. We have calculated a semiclassical
wave function of a doped graphene under the action of EM excitation and the
current through a molecular junction with graphene electrodes using
non-equilibrium Green functions. We have shown that using graphene electrodes
can essentially enhance currents evaluated at side-band energies $\sim
n\hbar\omega$ in molecular nanojunctions that is related to the modification
of the graphene gapless spectrum under the action of external EM field. We
have calculated the corresponding quasienergy spectrum that is accompanied
with opening the gap induced by intraband excitations.

If one shall use an electric field that is \textit{perpendicular} to the
graphene sheet, the field can excite $p$-polarized surface plasmons
propagating along the sheet with very high levels of spatial confinement and
large near-field enhancement \cite{Abajo11NL,Chen12Nature,Fei12Nature}.
Furthermore, surface plasmons in graphene have the advantage of being highly
tunable via electrostatic gating
\cite{Mak08PRL,Chen8Nature,Abajo11NL,Chen12Nature,Fei12Nature,Cox12PRB}. These
plasmon oscillations can enhance the dipole light-matter interaction in a
molecular bridge resulting in much more efficient control of photocurrent
related to the processes occurring in the molecular bridge under the action of
EM field polarized along the bridge
\cite{Kohler05,Kleinekathofer06EPL,Li07EPL,Li08SSP,Fainberg13CPL}. By this
means\ a side benifit of using doped graphene electrodes in molecular
nanojunctions is the \textit{polarization control} of the processes occurring
either in the graphene electrodes (if the electric field is parallel to the
graphene sheet) or in the molecular bridge (if the electric field is
perpendicular to the graphene sheet). Such selectivity may be achieved by
changing the polarization of an external EM field. This issue will be studied
in more detail elsewhere.

%\textbf{Acknowledgement}

\begin{acknowledgments}
The work has been supported in part by the US-Israel Binational Science
Foundation (grant No. 2008282). The author thanks A. Nitzan for useful discussion.
\end{acknowledgments}

%\bibliographystyle{apsrev4-1}
%\bibliography{ADV2,ADV2a,ADVCHEM,allref,Levinsky}

\begin{thebibliography}{10}%
\makeatletter
\providecommand \@ifxundefined [1]{%
 \ifx #1\undefined \expandafter \@firstoftwo
 \else \expandafter \@secondoftwo
\fi
}%
\providecommand \@ifnum [1]{%
 \ifnum #1\expandafter \@firstoftwo
 \else \expandafter \@secondoftwo
\fi
}%
\providecommand \enquote [1]{``#1''}%
\providecommand \bibnamefont  [1]{#1}%
\providecommand \bibfnamefont [1]{#1}%
\providecommand \citenamefont [1]{#1}%
\providecommand\href[0]{\@sanitize\@href}%
\providecommand\@href[1]{\endgroup\@@startlink{#1}\endgroup\@@href}%
\providecommand\@@href[1]{#1\@@endlink}%
\providecommand \@sanitize [0]{\begingroup\catcode`\&12\catcode`\#12\relax}%
\@ifxundefined \pdfoutput {\@firstoftwo}{%
 \@ifnum{\z@=\pdfoutput}{\@firstoftwo}{\@secondoftwo}%
}{%
 \providecommand\@@startlink[1]{\leavevmode}%
 \providecommand\@@endlink[0]{}%
}{%
 \providecommand\@@startlink[1]{%
  \leavevmode
  \pdfstartlink
   attr{/Border[0 0 1 ]/H/I/C[0 1 1]}%
   user{/Subtype/Link/A<</Type/Action/S/URI/URI(#1)>>}%
  \relax
 }%
 \providecommand\@@endlink[0]{\pdfendlink}%
}%
\providecommand \url  [0]{\begingroup\@sanitize \@url }%
\providecommand \@url [1]{\endgroup\@href {#1}{\urlprefix}}%
\providecommand \urlprefix [0]{URL }%
\providecommand \Eprint[0]{\href }%
\@ifxundefined \urlstyle {%
  \providecommand \doi [1]{doi:\discretionary{}{}{}#1}%
}{%
  \providecommand \doi [0]{doi:\discretionary{}{}{}\begingroup
  \urlstyle{rm}\Url }%
}%
\providecommand \doibase [0]{http://dx.doi.org/}%
\providecommand \Doi[1]{\href{\doibase#1}}%
\providecommand \bibAnnote [3]{%
  \BibitemShut{#1}%
  \begin{quotation}\noindent
    \textsc{Key:}\ #2\\\textsc{Annotation:}\ #3%
  \end{quotation}%
}%
\providecommand \bibAnnoteFile [2]{%
  \IfFileExists{#2}{\bibAnnote {#1} {#2} {\input{#2}}}{}%
}%
\providecommand \typeout [0]{\immediate \write \m@ne }%
\providecommand \selectlanguage [0]{\@gobble}%
\providecommand \bibinfo [0]{\@secondoftwo}%
\providecommand \bibfield [0]{\@secondoftwo}%
\providecommand \translation [1]{[#1]}%
\providecommand \BibitemOpen[0]{}%
\providecommand \bibitemStop [0]{}%
\providecommand \bibitemNoStop [0]{.\EOS\space}%
\providecommand \EOS [0]{\spacefactor3000\relax}%
\providecommand \BibitemShut [1]{\csname bibitem#1\endcsname}%
%</preamble>
\bibitem{Kohler05}%
  \BibitemOpen
  \bibfield{author}{%
  \bibinfo {author} {\bibfnamefont{S.}~\bibnamefont{Kohler}}, \bibinfo {author}
  {\bibfnamefont{J.}~\bibnamefont{Lehmann}},\ and\ \bibinfo {author}
  {\bibfnamefont{P.}~\bibnamefont{Hanggi}},\ }%
  \bibfield{journal}{%
  \bibinfo {journal} {Phys. Reports}\ }%
  \textbf{\bibinfo {volume} {406}},\ \bibinfo {pages} {379} (\bibinfo {year}
  {2005})%
  \bibAnnoteFile{NoStop}{Kohler05}%
\bibitem{Chen09}%
  \BibitemOpen
  \bibfield{author}{%
  \bibinfo {author} {\bibfnamefont{F.}~\bibnamefont{Chen}}\ and\ \bibinfo
  {author} {\bibfnamefont{N.~J.}\ \bibnamefont{Tao}},\ }%
  \bibfield{journal}{%
  \bibinfo {journal} {Accounts of Chemical Research}\ }%
  \textbf{\bibinfo {volume} {42}},\ \bibinfo {pages} {429} (\bibinfo {year}
  {2009})%
  \bibAnnoteFile{NoStop}{Chen09}%
\bibitem{Heath09}%
  \BibitemOpen
  \bibfield{author}{%
  \bibinfo {author} {\bibfnamefont{J.~R.}\ \bibnamefont{Heath}},\ }%
  \bibfield{journal}{%
  \bibinfo {journal} {Annual Review of Materials Research}\ }%
  \textbf{\bibinfo {volume} {39}},\ \bibinfo {pages} {1} (\bibinfo {year}
  {2009})%
  \bibAnnoteFile{NoStop}{Heath09}%
\bibitem{Park11PRB}%
  \BibitemOpen
  \bibfield{author}{%
  \bibinfo {author} {\bibfnamefont{T.-H.}\ \bibnamefont{Park}}\ and\ \bibinfo
  {author} {\bibfnamefont{M.}~\bibnamefont{Galperin}},\ }%
  \bibfield{journal}{%
  \bibinfo {journal} {Phys. Rev. B}\ }%
  \textbf{\bibinfo {volume} {84}},\ \bibinfo {pages} {075447} (\bibinfo {year}
  {2011})%
  \bibAnnoteFile{NoStop}{Park11PRB}%
\bibitem{Wang11PCCP}%
  \BibitemOpen
  \bibfield{author}{%
  \bibinfo {author} {\bibfnamefont{L.}~\bibnamefont{Wang}}\ and\ \bibinfo
  {author} {\bibfnamefont{V.}~\bibnamefont{May}},\ }%
  \bibfield{journal}{%
  \bibinfo {journal} {Phys. Chem. Chem. Phys}\ }%
  \textbf{\bibinfo {volume} {13}},\ \bibinfo {pages} {8755} (\bibinfo {year}
  {2011})%
  \bibAnnoteFile{NoStop}{Wang11PCCP}%
\bibitem{Fainberg_Galperin11PRB}%
  \BibitemOpen
  \bibfield{author}{%
  \bibinfo {author} {\bibfnamefont{B.~D.}\ \bibnamefont{Fainberg}}, \bibinfo
  {author} {\bibfnamefont{M.}~\bibnamefont{Sukharev}}, \bibinfo {author}
  {\bibfnamefont{T.-H.}\ \bibnamefont{Park}},\ and\ \bibinfo {author}
  {\bibfnamefont{M.}~\bibnamefont{Galperin}},\ }%
  \bibfield{journal}{%
  \bibinfo {journal} {Phys. Rev. B}\ }%
  \textbf{\bibinfo {volume} {83}},\ \bibinfo {pages} {205425} (\bibinfo {year}
  {2011})%
  \bibAnnoteFile{NoStop}{Fainberg_Galperin11PRB}%
\bibitem{Haertle10JCP}%
  \BibitemOpen
  \bibfield{author}{%
  \bibinfo {author} {\bibfnamefont{R.}~\bibnamefont{Haertle}}, \bibinfo
  {author} {\bibfnamefont{R.}~\bibnamefont{Volkovich}},\ and\ \bibinfo {author}
  {\bibfnamefont{M.}~\bibnamefont{Thoss}},\ }%
  \bibfield{journal}{%
  \bibinfo {journal} {J. Chem. Phys.}\ }%
  \textbf{\bibinfo {volume} {133}},\ \bibinfo {pages} {081102} (\bibinfo {year}
  {2010})%
  \bibAnnoteFile{NoStop}{Haertle10JCP}%
\bibitem{Reuter08PRL}%
  \BibitemOpen
  \bibfield{author}{%
  \bibinfo {author} {\bibfnamefont{M.~G.}\ \bibnamefont{Reuter}}, \bibinfo
  {author} {\bibfnamefont{M.}~\bibnamefont{Sukharev}},\ and\ \bibinfo {author}
  {\bibfnamefont{T.}~\bibnamefont{Seideman}},\ }%
  \bibfield{journal}{%
  \bibinfo {journal} {Phys. Rev. Lett.}\ }%
  \textbf{\bibinfo {volume} {101}},\ \bibinfo {pages} {208303} (\bibinfo {year}
  {2008})%
  \bibAnnoteFile{NoStop}{Reuter08PRL}%
\bibitem{Li08SSP}%
  \BibitemOpen
  \bibfield{author}{%
  \bibinfo {author} {\bibfnamefont{G.}~\bibnamefont{Li}}, \bibinfo {author}
  {\bibfnamefont{M.}~\bibnamefont{Schreiber}},\ and\ \bibinfo {author}
  {\bibfnamefont{U.}~\bibnamefont{Kleinekathoefer}},\ }%
  \bibfield{journal}{%
  \bibinfo {journal} {Physica Status Solidi B-Basic Solid State Physics}\ }%
  \textbf{\bibinfo {volume} {245}},\ \bibinfo {pages} {2720} (\bibinfo {year}
  {2008})%
  \bibAnnoteFile{NoStop}{Li08SSP}%
\bibitem{Thanopulos08Nanotech}%
  \BibitemOpen
  \bibfield{author}{%
  \bibinfo {author} {\bibfnamefont{I.}~\bibnamefont{Thanopulos}}, \bibinfo
  {author} {\bibfnamefont{E.}~\bibnamefont{Paspalakis}},\ and\ \bibinfo
  {author} {\bibfnamefont{V.}~\bibnamefont{Yannopapas}},\ }%
  \bibfield{journal}{%
  \bibinfo {journal} {Nanotechnology}\ }%
  \textbf{\bibinfo {volume} {19}},\ \bibinfo {pages} {445202} (\bibinfo {year}
  {2008})%
  \bibAnnoteFile{NoStop}{Thanopulos08Nanotech}%
\bibitem{Prociuk08PRB}%
  \BibitemOpen
  \bibfield{author}{%
  \bibinfo {author} {\bibfnamefont{A.}~\bibnamefont{Prociuk}}\ and\ \bibinfo
  {author} {\bibfnamefont{B.~D.}\ \bibnamefont{Dunietz}},\ }%
  \bibfield{journal}{%
  \bibinfo {journal} {Phys. Rev. B}\ }%
  \textbf{\bibinfo {volume} {78}},\ \bibinfo {pages} {165112} (\bibinfo {year}
  {2008})%
  \bibAnnoteFile{NoStop}{Prociuk08PRB}%
\bibitem{Li08NJP}%
  \BibitemOpen
  \bibfield{author}{%
  \bibinfo {author} {\bibfnamefont{G.}~\bibnamefont{Li}}, \bibinfo {author}
  {\bibfnamefont{M.}~\bibnamefont{Schreiber}},\ and\ \bibinfo {author}
  {\bibfnamefont{U.}~\bibnamefont{Kleinekathoefer}},\ }%
  \bibfield{journal}{%
  \bibinfo {journal} {New J. Phys.}\ }%
  \textbf{\bibinfo {volume} {10}},\ \bibinfo {pages} {085005} (\bibinfo {year}
  {2008})%
  \bibAnnoteFile{NoStop}{Li08NJP}%
\bibitem{Galperin08JCP}%
  \BibitemOpen
  \bibfield{author}{%
  \bibinfo {author} {\bibfnamefont{M.}~\bibnamefont{Galperin}}\ and\ \bibinfo
  {author} {\bibfnamefont{S.}~\bibnamefont{Tretiak}},\ }%
  \bibfield{journal}{%
  \bibinfo {journal} {J. Chem. Phys.}\ }%
  \textbf{\bibinfo {volume} {128}},\ \bibinfo {pages} {124705} (\bibinfo {year}
  {2008})%
  \bibAnnoteFile{NoStop}{Galperin08JCP}%
\bibitem{Li07EPL}%
  \BibitemOpen
  \bibfield{author}{%
  \bibinfo {author} {\bibfnamefont{U.}~\bibnamefont{Kleinekathofer}}, \bibinfo
  {author} {\bibfnamefont{G.}~\bibnamefont{Li}}, \bibinfo {author}
  {\bibfnamefont{S.}~\bibnamefont{Welack}},\ and\ \bibinfo {author}
  {\bibfnamefont{M.}~\bibnamefont{Schreiber}},\ }%
  \bibfield{journal}{%
  \bibinfo {journal} {Europhys. Letters}\ }%
  \textbf{\bibinfo {volume} {79}},\ \bibinfo {pages} {27006} (\bibinfo {year}
  {2007})%
  \bibAnnoteFile{NoStop}{Li07EPL}%
\bibitem{Fai07PRB}%
  \BibitemOpen
  \bibfield{author}{%
  \bibinfo {author} {\bibfnamefont{B.~D.}\ \bibnamefont{Fainberg}}, \bibinfo
  {author} {\bibfnamefont{M.}~\bibnamefont{Jouravlev}},\ and\ \bibinfo {author}
  {\bibfnamefont{A.}~\bibnamefont{Nitzan}},\ }%
  \bibfield{journal}{%
  \bibinfo {journal} {Phys. Rev. B}\ }%
  \textbf{\bibinfo {volume} {76}},\ \bibinfo {pages} {245329} (\bibinfo {year}
  {2007})%
  \bibAnnoteFile{NoStop}{Fai07PRB}%
\bibitem{Platero04}%
  \BibitemOpen
  \bibfield{author}{%
  \bibinfo {author} {\bibfnamefont{G.}~\bibnamefont{Platero}}\ and\ \bibinfo
  {author} {\bibfnamefont{R.}~\bibnamefont{Aguado}},\ }%
  \bibfield{journal}{%
  \bibinfo {journal} {Phys. Reports}\ }%
  \textbf{\bibinfo {volume} {395}},\ \bibinfo {pages} {1} (\bibinfo {year}
  {2004})%
  \bibAnnoteFile{NoStop}{Platero04}%
\bibitem{Dayem_Martin62PRL}%
  \BibitemOpen
  \bibfield{author}{%
  \bibinfo {author} {\bibfnamefont{A.~H.~L.}\ \bibnamefont{Dayem}}\ and\
  \bibinfo {author} {\bibfnamefont{R.~J.}\ \bibnamefont{Martin}},\ }%
  \bibfield{journal}{%
  \bibinfo {journal} {Phys. Rev. Lett.}\ }%
  \textbf{\bibinfo {volume} {8}},\ \bibinfo {pages} {246} (\bibinfo {year}
  {1962})%
  \bibAnnoteFile{NoStop}{Dayem_Martin62PRL}%
\bibitem{Tien_Gordon63}%
  \BibitemOpen
  \bibfield{author}{%
  \bibinfo {author} {\bibfnamefont{P.~K.}\ \bibnamefont{Tien}}\ and\ \bibinfo
  {author} {\bibfnamefont{J.~P.}\ \bibnamefont{Gordon}},\ }%
  \bibfield{journal}{%
  \bibinfo {journal} {Phys. Rev.}\ }%
  \textbf{\bibinfo {volume} {129}},\ \bibinfo {pages} {647} (\bibinfo {year}
  {1963})%
  \bibAnnoteFile{NoStop}{Tien_Gordon63}%
\bibitem{Gri98}%
  \BibitemOpen
  \bibfield{author}{%
  \bibinfo {author} {\bibfnamefont{M.}~\bibnamefont{Grifoni}}\ and\ \bibinfo
  {author} {\bibfnamefont{P.}~\bibnamefont{Hanggi}},\ }%
  \bibfield{journal}{%
  \bibinfo {journal} {Phys. Reports}\ }%
  \textbf{\bibinfo {volume} {304}},\ \bibinfo {pages} {229} (\bibinfo {year}
  {1998})%
  \bibAnnoteFile{NoStop}{Gri98}%
\bibitem{Kleinekathofer06EPL}%
  \BibitemOpen
  \bibfield{author}{%
  \bibinfo {author} {\bibfnamefont{U.}~\bibnamefont{Kleinekathofer}}, \bibinfo
  {author} {\bibfnamefont{G.}~\bibnamefont{Li}}, \bibinfo {author}
  {\bibfnamefont{S.}~\bibnamefont{Welack}},\ and\ \bibinfo {author}
  {\bibfnamefont{M.}~\bibnamefont{Schreiber}},\ }%
  \bibfield{journal}{%
  \bibinfo {journal} {Europhys. Letters}\ }%
  \textbf{\bibinfo {volume} {75}},\ \bibinfo {pages} {139} (\bibinfo {year}
  {2006})%
  \bibAnnoteFile{NoStop}{Kleinekathofer06EPL}%
\bibitem{Novoselov09RMP}%
  \BibitemOpen
  \bibfield{author}{%
  \bibinfo {author} {\bibfnamefont{A.~H.~C.}\ \bibnamefont{Neto}}, \bibinfo
  {author} {\bibfnamefont{F.}~\bibnamefont{Guinea}}, \bibinfo {author}
  {\bibfnamefont{N.~M.~R.}\ \bibnamefont{Peres}}, \bibinfo {author}
  {\bibfnamefont{K.~S.}\ \bibnamefont{Novoselov}},\ and\ \bibinfo {author}
  {\bibfnamefont{A.~K.}\ \bibnamefont{Geim}},\ }%
  \bibfield{journal}{%
  \bibinfo {journal} {Rev. Mod. Phys.}\ }%
  \textbf{\bibinfo {volume} {81}},\ \bibinfo {pages} {109} (\bibinfo {year}
  {2009})%
  \bibAnnoteFile{NoStop}{Novoselov09RMP}%
\bibitem{Trauzettel07PRB}%
  \BibitemOpen
  \bibfield{author}{%
  \bibinfo {author} {\bibfnamefont{B.}~\bibnamefont{Trauzettel}}, \bibinfo
  {author} {\bibfnamefont{Y.~M.}\ \bibnamefont{Blanter}},\ and\ \bibinfo
  {author} {\bibfnamefont{A.~F.}\ \bibnamefont{Morpurgo}},\ }%
  \bibfield{journal}{%
  \bibinfo {journal} {Phys. Rev. B}\ }%
  \textbf{\bibinfo {volume} {75}},\ \bibinfo {pages} {035305} (\bibinfo {year}
  {2007})%
  \bibAnnoteFile{NoStop}{Trauzettel07PRB}%
\bibitem{Efetov08PRB}%
  \BibitemOpen
  \bibfield{author}{%
  \bibinfo {author} {\bibfnamefont{S.~V.}\ \bibnamefont{Syzranov}}, \bibinfo
  {author} {\bibfnamefont{M.~V.}\ \bibnamefont{Fistul}},\ and\ \bibinfo
  {author} {\bibfnamefont{K.~B.}\ \bibnamefont{Efetov}},\ }%
  \bibfield{journal}{%
  \bibinfo {journal} {Phys. Rev. B}\ }%
  \textbf{\bibinfo {volume} {78}},\ \bibinfo {pages} {045407} (\bibinfo {year}
  {2008})%
  \bibAnnoteFile{NoStop}{Efetov08PRB}%
\bibitem{Yang_graphene_junctions10JCP}%
  \BibitemOpen
  \bibfield{author}{%
  \bibinfo {author} {\bibfnamefont{X.}~\bibnamefont{Zheng}}, \bibinfo {author}
  {\bibfnamefont{S.-H.}\ \bibnamefont{Ke}},\ and\ \bibinfo {author}
  {\bibfnamefont{W.}~\bibnamefont{Yang}},\ }%
  \bibfield{journal}{%
  \bibinfo {journal} {J. of Chem. Phys}\ }%
  \textbf{\bibinfo {volume} {132}},\ \bibinfo {pages} {114703} (\bibinfo {year}
  {2010})%
  \bibAnnoteFile{NoStop}{Yang_graphene_junctions10JCP}%
\bibitem{Mikhailov07EPL}%
  \BibitemOpen
  \bibfield{author}{%
  \bibinfo {author} {\bibfnamefont{S.~A.}\ \bibnamefont{Mikhailov}},\ }%
  \bibfield{journal}{%
  \bibinfo {journal} {Europhys. Letters}\ }%
  \textbf{\bibinfo {volume} {79}},\ \bibinfo {pages} {27002} (\bibinfo {year}
  {2007})%
  \bibAnnoteFile{NoStop}{Mikhailov07EPL}%
\bibitem{Fainberg13CPL}%
  \BibitemOpen
  \bibfield{author}{%
  \bibinfo {author} {\bibfnamefont{B.~D.}\ \bibnamefont{Fainberg}}\ and\
  \bibinfo {author} {\bibfnamefont{T.}~\bibnamefont{Seideman}},\ }%
  \bibfield{journal}{%
  \bibinfo {journal} {Chem. Phys. Lett.}\ }%
  \textbf{\bibinfo {volume} {576}} (\bibinfo {year} {2013}),\ \bibinfo {note}
  {[Frontiers Article]}%
  \bibAnnoteFile{NoStop}{Fainberg13CPL}%
\bibitem{Abajo11NL}%
  \BibitemOpen
  \bibfield{author}{%
  \bibinfo {author} {\bibfnamefont{F.~H.~L.}\ \bibnamefont{Koppens}}, \bibinfo
  {author} {\bibfnamefont{D.~E.}\ \bibnamefont{Chang}},\ and\ \bibinfo {author}
  {\bibfnamefont{F.~J.~G.}\ \bibnamefont{de~Abajo}},\ }%
  \bibfield{journal}{%
  \bibinfo {journal} {Nano Letters}\ }%
  \textbf{\bibinfo {volume} {11}},\ \bibinfo {pages} {3370} (\bibinfo {year}
  {2011})%
  \bibAnnoteFile{NoStop}{Abajo11NL}%
\bibitem{Chen12Nature}%
  \BibitemOpen
  \bibfield{author}{%
  \bibinfo {author} {\bibfnamefont{J.}~\bibnamefont{Chen}}, \bibinfo {author}
  {\bibfnamefont{M.}~\bibnamefont{Badioli}}, \bibinfo {author}
  {\bibfnamefont{P.}~\bibnamefont{Alonso-Gonzalez}}, \bibinfo {author}
  {\bibfnamefont{S.}~\bibnamefont{Thongrattanasiri}}, \bibinfo {author}
  {\bibfnamefont{F.}~\bibnamefont{Huth}}, \bibinfo {author}
  {\bibfnamefont{J.}~\bibnamefont{Osmond}}, \bibinfo {author}
  {\bibfnamefont{M.}~\bibnamefont{Spasenovic}}, \bibinfo {author}
  {\bibfnamefont{A.}~\bibnamefont{Centeno}}, \bibinfo {author}
  {\bibfnamefont{A.}~\bibnamefont{Pesquera}}, \bibinfo {author}
  {\bibfnamefont{P.}~\bibnamefont{Godignon}}, \bibinfo {author}
  {\bibfnamefont{A.~Z.}\ \bibnamefont{Elorza}}, \bibinfo {author}
  {\bibfnamefont{N.}~\bibnamefont{Camara}}, \bibinfo {author}
  {\bibfnamefont{F.~J.~G.}\ \bibnamefont{de~Abajo}}, \bibinfo {author}
  {\bibfnamefont{R.}~\bibnamefont{Hillenbrand}},\ and\ \bibinfo {author}
  {\bibfnamefont{F.~H.~L.}\ \bibnamefont{Koppens}},\ }%
  \bibfield{journal}{%
  \bibinfo {journal} {Nature}\ }%
  \textbf{\bibinfo {volume} {487}},\ \bibinfo {pages} {77} (\bibinfo {year}
  {2012})%
  \bibAnnoteFile{NoStop}{Chen12Nature}%
\bibitem{Fei12Nature}%
  \BibitemOpen
  \bibfield{author}{%
  \bibinfo {author} {\bibfnamefont{Z.}~\bibnamefont{Fei}}, \bibinfo {author}
  {\bibfnamefont{A.~S.}\ \bibnamefont{Rodin}}, \bibinfo {author}
  {\bibfnamefont{G.~O.}\ \bibnamefont{Andreev}}, \bibinfo {author}
  {\bibfnamefont{W.}~\bibnamefont{Bao}}, \bibinfo {author}
  {\bibfnamefont{A.~S.}\ \bibnamefont{McLeod}}, \bibinfo {author}
  {\bibfnamefont{M.}~\bibnamefont{Wagner}}, \bibinfo {author}
  {\bibfnamefont{L.~M.}\ \bibnamefont{Zhang}}, \bibinfo {author}
  {\bibfnamefont{Z.}~\bibnamefont{Zhao}}, \bibinfo {author}
  {\bibfnamefont{M.}~\bibnamefont{Thiemens}}, \bibinfo {author}
  {\bibfnamefont{G.}~\bibnamefont{Dominguez}}, \bibinfo {author}
  {\bibfnamefont{M.~M.}\ \bibnamefont{Fogler}}, \bibinfo {author}
  {\bibfnamefont{A.~H.~C.}\ \bibnamefont{Neto}}, \bibinfo {author}
  {\bibfnamefont{C.~N.}\ \bibnamefont{Lau}}, \bibinfo {author}
  {\bibfnamefont{F.}~\bibnamefont{Keilmann}},\ and\ \bibinfo {author}
  {\bibfnamefont{D.~N.}\ \bibnamefont{Basov}},\ }%
  \bibfield{journal}{%
  \bibinfo {journal} {Nature}\ }%
  \textbf{\bibinfo {volume} {487}},\ \bibinfo {pages} {82} (\bibinfo {year}
  {2012})%
  \bibAnnoteFile{NoStop}{Fei12Nature}%
\bibitem{Mak08PRL}%
  \BibitemOpen
  \bibfield{author}{%
  \bibinfo {author} {\bibfnamefont{K.~F.}\ \bibnamefont{Mak}}, \bibinfo
  {author} {\bibfnamefont{M.~Y.}\ \bibnamefont{Sfeir}}, \bibinfo {author}
  {\bibfnamefont{Y.}~\bibnamefont{Wu}}, \bibinfo {author}
  {\bibfnamefont{C.~H.}\ \bibnamefont{Lui}}, \bibinfo {author}
  {\bibfnamefont{J.~A.}\ \bibnamefont{Misewich}},\ and\ \bibinfo {author}
  {\bibfnamefont{T.~F.}\ \bibnamefont{Heinz}},\ }%
  \bibfield{journal}{%
  \bibinfo {journal} {Phys. Rev. Lett.}\ }%
  \textbf{\bibinfo {volume} {101}},\ \bibinfo {pages} {196405} (\bibinfo {year}
  {2008})%
  \bibAnnoteFile{NoStop}{Mak08PRL}%
\bibitem{Chen8Nature}%
  \BibitemOpen
  \bibfield{author}{%
  \bibinfo {author} {\bibfnamefont{C.-F.}\ \bibnamefont{Chen}}, \bibinfo
  {author} {\bibfnamefont{C.-H.}\ \bibnamefont{Park}}, \bibinfo {author}
  {\bibfnamefont{B.~W.}\ \bibnamefont{Boudouris}}, \bibinfo {author}
  {\bibfnamefont{J.}~\bibnamefont{Horng}}, \bibinfo {author}
  {\bibfnamefont{B.}~\bibnamefont{Geng}}, \bibinfo {author}
  {\bibfnamefont{C.}~\bibnamefont{Girit}}, \bibinfo {author}
  {\bibfnamefont{A.}~\bibnamefont{Zettl}}, \bibinfo {author}
  {\bibfnamefont{M.~F.}\ \bibnamefont{Crommie}}, \bibinfo {author}
  {\bibfnamefont{R.~A.}\ \bibnamefont{Segalman}}, \bibinfo {author}
  {\bibfnamefont{S.~G.}\ \bibnamefont{Louie}},\ and\ \bibinfo {author}
  {\bibfnamefont{F.}~\bibnamefont{Wang}},\ }%
  \bibfield{journal}{%
  \bibinfo {journal} {Nature}\ }%
  \textbf{\bibinfo {volume} {471}},\ \bibinfo {pages} {617} (\bibinfo {year}
  {2011})%
  \bibAnnoteFile{NoStop}{Chen8Nature}%
\bibitem{Pauli32}%
  \BibitemOpen
  \bibfield{author}{%
  \bibinfo {author} {\bibfnamefont{W.}~\bibnamefont{Pauli}},\ }%
  \bibfield{journal}{%
  \bibinfo {journal} {Helv. Phys. Acta}\ }%
  \textbf{\bibinfo {volume} {5}},\ \bibinfo {pages} {179} (\bibinfo {year}
  {1932})%
  \bibAnnoteFile{NoStop}{Pauli32}%
\bibitem{akhiezer-berestetskii69}%
  \BibitemOpen
  \bibfield{author}{%
  \bibinfo {author} {\bibfnamefont{A.~I.}\ \bibnamefont{Akhiezer}}\ and\
  \bibinfo {author} {\bibfnamefont{V.~B.}\ \bibnamefont{Berestetskii}},\ }%
  \emph{\bibinfo {title} {Quantum electrodynamics}}\ (\bibinfo {publisher}
  {Nauka},\ \bibinfo {address} {Moskow},\ \bibinfo {year} {1969})\ \bibinfo
  {note} {in Russian}%
  \bibAnnoteFile{NoStop}{akhiezer-berestetskii69}%
\bibitem{berestetskii-lifshitz99}%
  \BibitemOpen
  \bibfield{author}{%
  \bibinfo {author} {\bibfnamefont{V.~B.}\ \bibnamefont{Berestetskii}},
  \bibinfo {author} {\bibfnamefont{E.~M.}\ \bibnamefont{Lifshitz}},\ and\
  \bibinfo {author} {\bibfnamefont{L.~P.}\ \bibnamefont{Pitaevskii}},\ }%
  \emph{\bibinfo {title} {Quantum electrodynamics}}\ (\bibinfo {publisher}
  {Butterworth-Heinemann},\ \bibinfo {address} {Oxford},\ \bibinfo {year}
  {1999})%
  \bibAnnoteFile{NoStop}{berestetskii-lifshitz99}%
\bibitem{landau-lifshitz.65}%
  \BibitemOpen
  \bibfield{author}{%
  \bibinfo {author} {\bibfnamefont{L.~D.}\ \bibnamefont{Landau}}\ and\ \bibinfo
  {author} {\bibfnamefont{E.~M.}\ \bibnamefont{Lifshitz}},\ }%
  \emph{\bibinfo {title} {Quantum mechanics non-relativistic theory}}\
  (\bibinfo {publisher} {Pergamon Press},\ \bibinfo {address} {New York},\
  \bibinfo {year} {1965})%
  \bibAnnoteFile{NoStop}{landau-lifshitz.65}%
\bibitem{haug-jauho.96}%
  \BibitemOpen
  \bibfield{author}{%
  \bibinfo {author} {\bibfnamefont{H.}~\bibnamefont{Haug}}\ and\ \bibinfo
  {author} {\bibfnamefont{A.~P.}\ \bibnamefont{Jauho}},\ }%
  \emph{\bibinfo {title} {Quantum Kinetics in Transportand Optics of
  Semiconductors}}\ (\bibinfo {publisher} {Springer},\ \bibinfo {address}
  {Berlin},\ \bibinfo {year} {1996})%
  \bibAnnoteFile{NoStop}{haug-jauho.96}%
\bibitem{Abr64}%
  \BibitemOpen
  \bibfield{author}{%
  \bibinfo {author} {\bibfnamefont{M.}~\bibnamefont{Abramowitz}}\ and\ \bibinfo
  {author} {\bibfnamefont{I.}~\bibnamefont{Stegun}},\ }%
  \emph{\bibinfo {title} {Handbook on Mathematical Functions}}\ (\bibinfo
  {publisher} {Dover},\ \bibinfo {address} {New York},\ \bibinfo {year}
  {1964})%
  \bibAnnoteFile{NoStop}{Abr64}%
\bibitem{Li_Fai12Nano_Let}%
  \BibitemOpen
  \bibfield{author}{%
  \bibinfo {author} {\bibfnamefont{G.}~\bibnamefont{Li}}, \bibinfo {author}
  {\bibfnamefont{M.~S.}\ \bibnamefont{Shishodia}}, \bibinfo {author}
  {\bibfnamefont{B.~D.}\ \bibnamefont{Fainberg}}, \bibinfo {author}
  {\bibfnamefont{B.}~\bibnamefont{Apter}}, \bibinfo {author}
  {\bibfnamefont{M.}~\bibnamefont{Oren}}, \bibinfo {author}
  {\bibfnamefont{A.}~\bibnamefont{Nitzan}},\ and\ \bibinfo {author}
  {\bibfnamefont{M.}~\bibnamefont{Ratner}},\ }%
  \bibfield{journal}{%
  \bibinfo {journal} {Nano Letters}\ }%
  \textbf{\bibinfo {volume} {12}},\ \bibinfo {pages} {2228} (\bibinfo {year}
  {2012})%
  \bibAnnoteFile{NoStop}{Li_Fai12Nano_Let}%
\bibitem{Oka_Aoki09PRB}%
  \BibitemOpen
  \bibfield{author}{%
  \bibinfo {author} {\bibfnamefont{T.}~\bibnamefont{Oka}}\ and\ \bibinfo
  {author} {\bibfnamefont{H.}~\bibnamefont{Aoki}},\ }%
  \bibfield{journal}{%
  \bibinfo {journal} {Phys. Rev. B}\ }%
  \textbf{\bibinfo {volume} {79}},\ \bibinfo {pages} {081406} (\bibinfo {year}
  {2009})%
  \bibAnnoteFile{NoStop}{Oka_Aoki09PRB}%
\bibitem{Cox12PRB}%
  \BibitemOpen
  \bibfield{author}{%
  \bibinfo {author} {\bibfnamefont{J.~D.}\ \bibnamefont{Cox}}, \bibinfo
  {author} {\bibfnamefont{M.~R.}\ \bibnamefont{Singh}}, \bibinfo {author}
  {\bibfnamefont{G.}~\bibnamefont{Gumbs}}, \bibinfo {author}
  {\bibfnamefont{M.~A.}\ \bibnamefont{Anton}},\ and\ \bibinfo {author}
  {\bibfnamefont{F.}~\bibnamefont{Carreno}},\ }%
  \bibfield{journal}{%
  \bibinfo {journal} {Phys. Rev. B}\ }%
  \textbf{\bibinfo {volume} {86}},\ \bibinfo {pages} {125452} (\bibinfo {year}
  {2012})%
  \bibAnnoteFile{NoStop}{Cox12PRB}%
\end{thebibliography}
%

\end{document}